\documentclass{llncs}
\usepackage{amsmath,amscd,amsbsy,amssymb,latexsym,url,bm}
\usepackage{epsfig,epstopdf,graphicx,subfigure}
\usepackage{enumitem,balance,mathtools}
\usepackage{wrapfig}
\usepackage{epsfig}
\usepackage{amssymb}
\usepackage{amsmath}
\usepackage{amsfonts}
\usepackage{mathrsfs, euscript}
\usepackage[usenames]{color}
\usepackage{multicol}

\usepackage{colortbl}
\usepackage[table]{xcolor}
\usepackage{algorithm}
\usepackage{algorithmic}
\usepackage{authblk}
\usepackage[font={it}]{caption}
\usepackage[algo2e]{algorithm2e}
\usepackage{amssymb}
\usepackage[symbol]{footmisc}

\usepackage[margin=1.5in]{geometry}
\newtheorem{def:def}{Definition}
\newtheorem{thm:thm}{Theorem}
\newtheorem{thm:lm}{Lemma}
\usepackage{ifpdf}
\ifpdf
\else
\fi

\DeclareMathOperator*{\argmax}{arg \, max}

\DeclareMathOperator*{\cov}{Cov}
\newcommand{\bs}{\boldsymbol}

\newcommand{\bl}{\backslash}

\begin{document}

\title{A Theoretical Analysis of Two-Stage Recommendation
  for Cold-Start Collaborative Filtering}

\author[]{Xiaoxue Zhao, Jun Wang}
\institute{University College London, London, United Kingdom\\
\email{\{x.zhao,j.wang\}@cs.ucl.ac.uk}}


\renewcommand\Authands{ and }

\maketitle
\abstract { In this paper, we present a theoretical framework for
  tackling the cold-start collaborative filtering problem, where
  unknown targets (items or users) keep coming to the system, and
  there is a limited number of resources (users or items) that can be
  allocated and related to them.  The solution requires a trade-off
  between exploitation and exploration as with the limited
  recommendation opportunities, we need to, on one hand, allocate the
  most relevant resources right away, but, on the other hand, it is also necessary
  to allocate resources that are useful for learning the
  target's properties in order to recommend more relevant ones in the
  future. In this paper, we study a simple two-stage recommendation
  combining a sequential and a batch solution together.  We first
  model the problem with the partially observable Markov decision
  process (POMDP) and provide an exact solution. Then, through an
  in-depth analysis over the POMDP value iteration solution, we
  identify that an exact solution can be abstracted as selecting
  resources that are not only highly relevant to the target according
  to the initial-stage information, but also highly correlated, either
  positively or negatively, with
  other \textit{potential} resources for the next stage. With this
  finding, we propose an approximate solution to ease the
  intractability of the exact solution. Our initial results on
  synthetic data and the Movie Lens 100K dataset confirm the
  performance gains of our theoretical development and analysis.}

\vspace{-5pt}
\section{Introduction}
\label{sec:intro}

For approximately the last two decades, information retrieval has
fundamentally transformed the way in which people seek and work with
information. Roughly speaking, there are two types of information
retrieval (IR) systems \cite{belkin1992information}. On one hand, we
have \emph{ad hoc} information retrieval, e.g., web search
\cite{page1999pagerank}, which deals with a relatively fixed
collection of information items (webpages, documents, images, product
descriptions etc.) and dynamically changing users information
requests. On the other hand, there are information filtering systems, such as
content recommender systems, to address the situation where user profiles
(as information requests) stay relatively static while new information
items keep arriving. Nevertheless, in either case the fundamental
problem remains the same, which is how to compute and find the
\emph{match} between the information items and information requests
\cite{manning08}.

\vspace{10pt}
A more difficult scenario exists when there is little or no information about
the request.  For instance, in collaborative filtering (CF), it is
hard to initialise recommendations when no past ratings are available. Research has been focused on the user \emph{cold-start}
problem~\cite{good1999combining,schein2002methods}, such as adopting a
questionnaire stage
\cite{rashid2002getting,rashid2008learning,zhou2011functional}, or an
interactive procedure \cite{zhao2013interactive,good1999combining}.
For the item cold-start problem, the main focus has been put on
utilising content information
\cite{rubens2011active,gunawardana2008tied}, which lies outside of the scope
of CF, or experimental design \cite{anava2015budget}.

\vspace{10pt}
In our view, the cold-start problem can be regarded as a resource
allocation problem, because in a short period of time, the
number of recommendations (for a new item or to a new user) is
usually much smaller than the size of the available pool. Thus only
a small portion can be selected due to the limited resources. For
example, advertisements of a new item can only be sent to a
limited number of users, whereas a new user can only rate a limited
number of items after joining a web service.  Therefore, it is important
to utilise the limited recommendation resources wisely.

\vspace{10pt}

\begin{figure*}[t]
  \centering \vspace{-15pt} \subfigure[For a cold-start item.]{
	\includegraphics[width=3.7in]{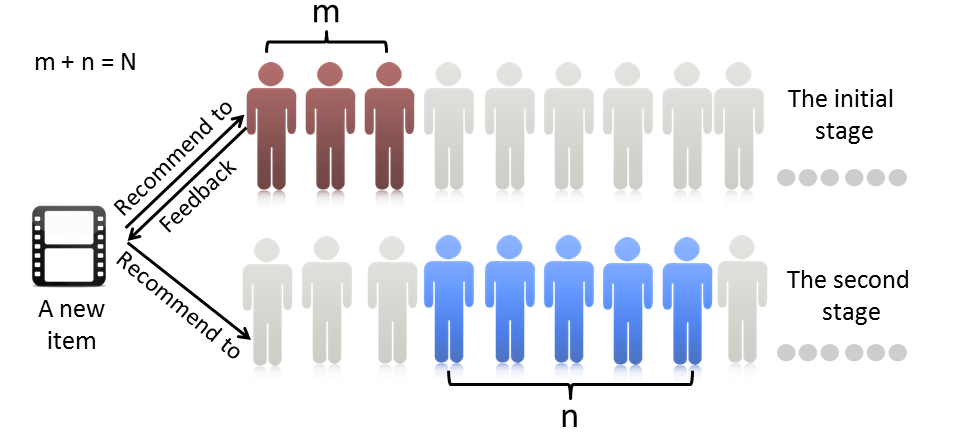}
	\label{fig:schematicmatic_item}
	}

\subfigure[For a cold-start user.]{
	\includegraphics[width=3.7in]{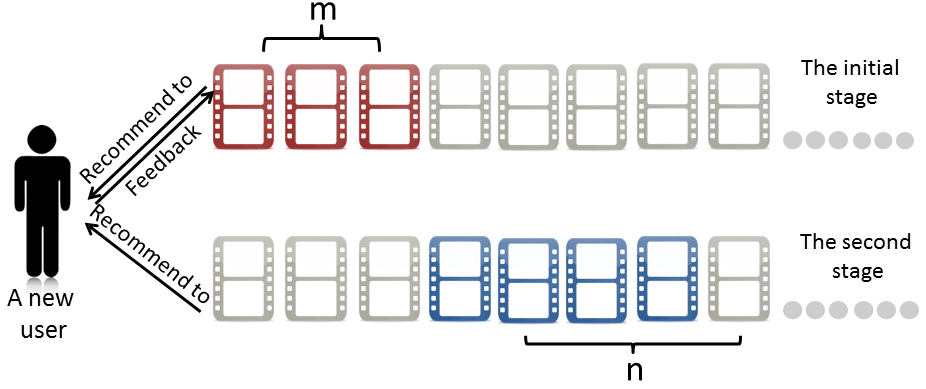}
	\label{fig:schematicmatic_user}
}
\vspace{-5pt}
\captionsetup{width=0.85\textwidth}
\caption{Schematic figures of the two-stage recommendation
 process for (a) a cold-start item, and (b) a cold-start user. The total $N$ resources are allocated in two stages. In the initial stage,
 $m$ users (items) are selected, with their feedback used to update the profile of the
 new item (user). Then another $n$ users (items) are selected in the second stage to exploit the updated profile. The target is
 to maximise the overall feedback over two stages.}
\label{fig:schematic}
\end{figure*}

In this paper, we formulate and analyse a simple yet practical
two-stage process to solve the recommendation allocation problem.
During the initial stage, we use a portion of recommendation
allocations to estimate the new item's (user's) model. After that, during
the second stage, we recommend the item (user) using the remaining
resources.
We argue that the goal of this process should
be to maximise the \textit{total} feedback over two stages,
which leads to
a trade-off between exploitation and exploration. This means that, with limited resources,
we should not separate the learning process from recommendation. Rather, recommendations should be
made right from the beginning while also intelligently accommodating the learning requirement.
 The proposed two-stage recommendation process is depicted in Figure
\ref{fig:schematic}. In CF, items and users are
usually modeled symmetrically \cite{wang2008unified,koren2009matrix},
and, as such, we will focus on the item cold-start problem as our
working example. However, all the analysis can be easily adapted to a
user cold-start scenario.

\vspace{10pt}
Dividing the recommendation process into two stages is simple yet
powerful as it combines both batch and feedback mechanisms
together. The motivations for our analysis on the setting are threefold. First, for the cold-start item, to learn its profile over
time, one way is to sequentially target the item to one user, observe
its feedback, update the item profile model, and find another user
with the updated model, in an interactive manner similar to
\cite{zhao2013interactive}.  However, as users differ in their
response times, waiting one user's response before targeting to the
other is practically infeasible in many cases. Second, it may also be
computationally too expensive for the system to update whenever a new
rating is registered. A two-stage process, by contrast, enables the
system to act economically. Third, statistically analysing the
separated two-stage process also enables a clear understanding of
the trade-off between exploitation-exploration (EE) embedded in many
practical applications.

\vspace{10pt}
The two-stage setup also covers a variety of other applications. For
example, in IR, when a query is issued, the system shows two
subsequent pages to the user such that the second-page results can be
refined \cite{shen2005implicit,jin2013interactive}. And, in online
display advertising, for a new campaign, in order to understand which
part of users should be targeted, the advertiser can spend some budget
to show the ads to different users and collect their feedback (i.e.,
ad click or conversion), and then after the warming-up stage, leverage
the users' feedback and refine the target user groups for higher
advertising performance \cite{zhang2014optimal}.

\vspace{10pt}
We first formulate the two-stage recommendation with a
POMDP framework.  We then derive the exact solution for both a
correlated-user model (CU) and a matrix factorisation (MF) model,
along with a discussion on the link between them.  After that, we
present our theoretical finding, i.e., the users to choose in the
initial stage should be those not only highly relevant
according to the initial-stage information, but also able to potentially guide us to find users with high expected values in the next stage with updated information. This ability of guidance can be further abstracted as a strong correlation
between the initial-stage users and potential second-stage users, no matter positive or negative.
With this finding, we propose the
approximation method \textit{guided} exploitation-exploration
(GEE). We argue that, as our objective differs from that of an upper
confidence bound (UCB) or an active learning (AL) approach, our
proposed GEE algorithm is significantly different from them.  The
effectiveness of the proposed solution is confirmed by our experiments,
conducted using both synthetic data and a real dataset.

\vspace{10pt}
The rest of this paper is organised as follows. We formulate the problem and present its exact solution in Section~\ref{sec:methodology-twostage}. In Section~\ref{sec:approximation}, we present the proposed approximate solution GEE and  afterwards we discuss the related previous work in Section \ref{sec:related-work}. Our experimental results are reported in Section \ref{sec:exp-twostage}, and Section \ref{sec:con} summarises and concludes this paper.

\vspace{-5pt}

\section{The Two-Stage Model}
\label{sec:methodology-twostage}

In this section, we formulate CF into the
POMDP framework, which will lead us to the exact solution of our problem.
A POMDP models a Markov decision process where the true current state of the system
 is partially unobservable  \cite{kaelbling1998planning}. In the scenario of the item cold-start recommendation,
 the true state is each user's genuine (potential) preference as to the new item, which is unknown for the users having not rated it.
 To model the decision process,
we start with a correlated-user (CU) model as a probabilistic description of the
memory-based models in CF \cite{herlocker1999algorithmic,deshpande2004item}
and formulate it with POMDP. Then, we decompose the user-item rating matrix to
gain its formulation in the domain of MF. We provide,
for each model, the exact solution on how to select users optimally in order to collect  maximal overall feedback from the users over two stages.

%
%

\subsection{Correlated-User Model with POMDP}
\label{sec:mg_pomdp}

\begin{figure}[t]
 \centering
    \includegraphics[width=3.5in]{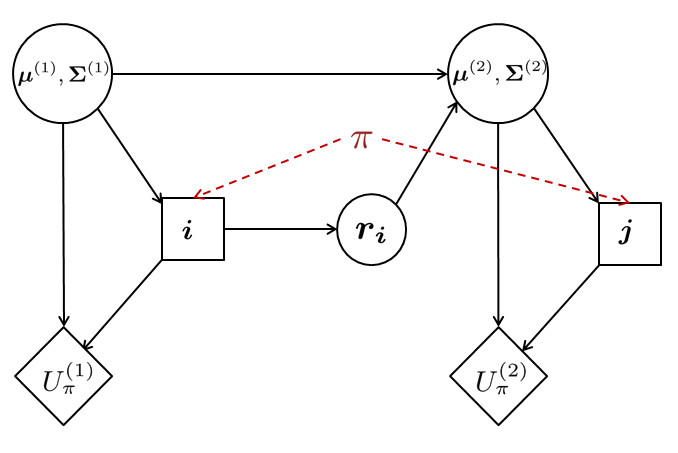}
    \captionsetup{width=0.85\textwidth}
 \caption{The two-stage CU-POMDP as illustrated by an influence diagram, with respect to
 the correlated-user model.
 Circular nodes are random variables and square nodes are the
 recommendation decision, and the rhombus nodes are the utility at each stage.  }
   \label{fig:mg_pomdp}
\end{figure}

The CU model with POMDP (CU-POMDP) is depicted in Figure~\ref{fig:mg_pomdp}.
Let us denote the available user pool as $\mathcal{U}$.
For each new item that joins the system,
the recommendation system should make the following decisions:
in the initial stage, choose an initial $m$ users
to start with, collect their feedback, and update the system's belief state;
and in the second stage, choose another $n$ users to exploit the information gained
from the initial stage. $N = m+n$ is the total number of users that the item is to be targeted to.
For the reader's convenience, we provide a list of key notations used in this paper in Table \ref{tab:twostage-notation}.
We consider  only one cold-start item, but
the scenario is similar if multiple cold-start items are present.

\begin{table}[t]
  \caption{Summary of key notations. }
  \label{tab:twostage-notation}
  \centering
\begin{tabular}{p{2.3cm} p{8.1cm}}
\hline
Notation & Description \\ \hline
$\mathcal{U}$ & The entire user set \\
$t \in \{1,2\}$ & The stage (timestep) of the process \\
$m,n$ & The number of users to select at the initial stage and the second stage respectively\\
$\bs{u}, \bs{v}$ & The users to choose in the initial stage and second stage respectively \\
$\bl \bs{u}$ & The users not selected in the initial stage, $\bl \bs{u} = \mathcal{U} \bl \bs{u}$\\
$\bs{R}$ & The preferences (a random vector) of all users over the item under consideration \\
$\bs{R_u}, \bs{R_v}$ & $\bs{R}$ partitioned by $\bs{u}$ and $\bs{v}$ respectively \\
$\bs{r_u},\bs{r_v}$ & Feedback from  $\bs{u}$ and $\bs{v}$ respectively \\
$\bs{\theta}^{(t)}, \mathbf{\Phi}^{(t)}, \mathbf{C}^{(t)}$ & The mean, covariance matrix, correlation matrix of $\bs{R}$ at time $t$ (CU model)\\
$\rho_{i,j}^{(t)}$ & Correlation between $u$ and $v$ at $t$\\
$\mathbf{P}$ & The matrix with each row as a user vector (MF model) \\
$\bs{q}$ & The target item's feature vector (MF model) \\
$\bs{\nu}^{(t)},\mathbf{\Psi}^{(t)}$ & The mean and covariance matrix of the item vector at time $t$ (MF model) \\
 $T$ & The sampling number \\
 \hline

\end{tabular}
\end{table}

\vspace{10pt}
Our goal is to find the  optimal policy that can maximise the expected total ratings over two stages.
To capture the relations between users' preferences, we model the
preferences of all users, denoted by $\bs{R}$, to follow a multivariate Gaussian distribution
\begin{align}
\label{eq:mu}
p^{(t)}(\bs{R}) \sim \mathcal{N}(\bs{\theta}^{(t)}, \mathbf{\Phi}^{(t)}), ~ t \in \{ 1,2\}
\end{align}
with its mean and covariance matrix as $\bs{\theta}^{(t)}$ and $\mathbf{\Phi}^{(t)}$. 
The distribution above is the system's belief over the true state $\bs{R}$ at each stage $t$, referred to as the belief state according to POMDP.
By recommending the item to users and receiving their feedback, the belief state evolves from $p^{(1)}(\bs{R})$ to
$p^{(2)}(\bs{R})$. Our problem is a POMDP  because the true
preferences $\bs{R}$ are unknown (or only partially known), but can be  modelled through a distribution.

%
%
%
%
%
%
%
%
%

\vspace{10pt}

This model is non-trivial because it has utilised all user-user correlations via a multivariate Gaussian model. To
 obtain the belief state for the initial stage, we can impose an i.i.d. assumption on
 the users' preferences on different items. As such, $\bs{\theta}^{(1)}$ can be estimated by the users' mean ratings,
and $\mathbf{\Phi}^{(1)}$ can be estimated by the user-user covariances on previously co-rated items.
To emphasise the role of user-user correlation, in the following, we also make use of the following representation
\begin{align}
\mathbf{\Phi}^{(1)}&= \text{Dg}[\bs{\Phi}^{(1)}]^{1/2}\mathbf{C}^{(1)}\text{Dg}[\bs{\Phi}^{(1)}]^{1/2} \nonumber \\ 
&= \text{diag}[\bs{\phi}^{(1)}]\mathbf{C}^{(1)}\text{diag}[\bs{\phi}^{(1)}]
\label{eq:corr}
\end{align}
where $\text{Dg}(\mathbf{\Phi}^{(1)})$ denotes the diagonal matrix with the same diagonal elements of $\mathbf{\Phi}^{(1)}$,
$\bs{\phi}^{(1)}$ denotes the vector formed by the users' standard deviations of ratings ($\bs{\phi}^{(1)} = \text{diag}[\text{Dg}^{1/2}(\mathbf{\Phi}^{(1)})]$)
,
and $\mathbf{C}^{(1)}$ is the correlation matrix whose element $\rho_{u,v}^{(1)}$ is the correlation
between user $u$ and user $v$.

 A policy $\pi$ is defined to make the decision at each stage on the basis of the available information:
\begin{align}
\bs{u} &= \pi(\bs{\theta}^{(1)}, \mathbf{\Phi}^{(1)}, \mathcal{U}), ~\text{and} \\
\bs{v} &= \pi(\bs{\theta}^{(2)}, \mathbf{\Phi}^{(2)}, \mathcal{U}\bl \bs{u}),
\end{align}
where we use vectors $\bs{u}$ and $\bs{v}$ to denote the user selection decisions for the two stages respectively
($|\bs{u}| = m$ and $|\bs{v}|=n$). Here we also use the constraint that the target item should not be recommended repeatedly to the same user. Therefore, the available user pool will be the remaining users $\mathcal{U}\bl \bs{u}$ for the second stage.
The total expected ratings
collected at each stage is the element-wise summation of the expected rating vector of each selection, which
we refer to as reward  $U_\pi^{(t)}$
\begin{align}
U_{\pi}^{(1)} = \mathbb{E}^{(1)}[\bs{1}^T \bs{R_u}],\\
U_{\pi}^{(2)} = \mathbb{E}^{(2)}[\bs{1}^T \bs{R_v}].
\end{align}

\vspace{10pt}
We use $\bs{R_u}$ ($\bs{R_v}$) to denote the random vector $\bs{R}$ partitioned by user selections $\bs{u}$ ($\bs{v}$). We will use the same partition rule throughout this paper.

\vspace{10pt}
The objective is to find a policy of selecting users such that the expected \textit{total} reward of the two stages are maximised
\begin{align}
\pi^\ast = \argmax_{\pi}\big( U_{\pi}^{(1)} + U_{\pi}^{(2)} \big).
\label{eq:object}
\end{align}

\subsubsection{\textbf{Belief Update}}

Let us consider the problem in a reverse order. Suppose the system has already
recommended the item to $m$ users in the initial stage and received feedback $\bs{r_u}$.
Given the feedback, the system can update
its belief state on the remaining users $\mathcal{U}\bl \bs{u}$ (simplified as $\bl \bs{u}$) by the conditional multivariate Gaussian distribution, conditioned on the observations
\begin{align}
\label{eq:cond_ri}
p^{(2)}(\bs{R}_{\bl \bs{u}}) &\sim \mathcal{N}(\bs{\theta}^{(2)}_{\bl
 \bs{u}}, \mathbf{\Phi}^{(2)}_{\bs{\bl u, \bl u}}),
 ~ \text{where}\\
\bs{\theta}^{(2)}_{\bl \bs{u}} &=\bs{\theta}^{(1)}_{\bl \bs{u}} + \mathbf{\Phi}^{(1)}_{\bs{\bl u,u}}[\mathbf{\Phi}^{(1)}_{\bs{u,u}}]^{-1}(\bs{r_u}-\bs{\theta}^{(1)}_{\bs{u}}) \label{eq:update_mu}\\
\mathbf{\Phi}^{(2)}_{\bs{\bl u, \bl u}} & = \mathbf{\Phi}^{(1)}_{\bs{\bl u,\bl u}} - \mathbf{\Phi}^{(1)}_{\bs{\bl u, u}} [\mathbf{\Phi}^{(1)}_{\bs{u,u}}]^{-1} \mathbf{\Phi}^{(1)}_{\bs{u,\bl u}}. \label{eq:update_sigma}
\end{align}

\vspace{10pt}
To gain insight with the view of correlated users, we reformulate the update functions with the correlation matrix $\mathbf{C}^{(1)}$ as follows. According to  Eq. (\ref{eq:corr}), we obtain
\begin{align}
\label{eq:sigmaii}
[\mathbf{\Phi}^{(1)}_{\bs{u,u}}]^{-1} &= \text{diag}[\bs{\phi_u}^{(1)}]^{-1} [ \mathbf{C}_{\bs{u,u}}^{(1)}]^{-1}\text{diag}[\bs{\phi_u}^{(1)}]^{-1},~\text{and} \\
[\mathbf{\Phi}^{(1)}_{\bs{\bl u,u}}] &= \text{diag}[\bs{\phi_{\bl u}}^{(1)}]  \mathbf{C}_{\bs{\bl u,u}}^{(1)}\text{diag}[\bs{\phi_u}^{(1)}].
\label{eq:sigmaji}
\end{align}

\vspace{10pt}
Substituting Eqs. (\ref{eq:sigmaji}) and (\ref{eq:sigmaii}) into (\ref{eq:update_mu}) we further get
\begin{align}
\label{eq:update_mu_corr}
\bs{\theta}^{(2)}_{\bl \bs{u}}&=\bs{\theta}^{(1)}_{\bl \bs{u}}+  \text{diag}[\bs{\phi_{\bl u}}^{(1)}]  \mathbf{C}_{\bs{\bl u,u}}^{(1)}[ \mathbf{C}_{\bs{u,u}}^{(1)}]^{-1}\text{diag}[\bs{\phi_u}^{(1)}]^{-1}(\bs{r_u}-\bs{\theta}^{(1)}_{\bs{u}})
\end{align}

\vspace{10pt}
Particularly, if we assume equal rating variance for all users,  and disregard the correlations among $\bs{u}$ such that $\mathbf{C}_{\bs{u,u}}^{(1)}$ becomes an identity matrix, then Eq. (\ref{eq:update_mu_corr}) reduces
to a weighted summation of the observed ratings centred by their prior expectations $\bs{r_u} - \bs{\theta_u}^{(1)}$, with the weights as the correlations between unobserved users and observed users
\begin{align}
\bs{\theta}^{(2)}_{\bl \bs{u}} =\bs{\theta}^{(1)}_{\bl \bs{u}} +  \mathbf{C}_{\bs{\bl u,u}}^{(1)}(\bs{r_u}-\bs{\theta}^{(1)}_{\bs{u}}).
\label{eq:update_mu_approx}
\end{align}

\vspace{10pt}
Eq. (\ref{eq:update_mu_approx}) looks very familiar to us because it simulates the popular memory-based (user-based) CF algorithm , which
takes the neighbours' ratings regarding the target item, centres them by the mean ratings of the neighbours, and estimates the target user's preference regarding this item as their weighted summation \cite{herlocker1999algorithmic}, where
Pearson correlation is commonly used to calculate the weights \cite{rubens2011active}.
We thus see the user-based recommendation heuristic as an approximation of our CU model.

\vspace{10pt}
From the above formula we can see that: (i) by observing users
$\bs{u}$ in the initial stage, the expectations of unobserved users
are also updated; (ii) the covariances (correlations) between observed
and unobserved users act as the bridge through which feedback from
selected users can update our belief regarding other users.

\subsubsection{\textbf{Exact Solution}}

\vspace{-5pt}
\label{sec:mg_value_iteration}
To obtain the exact solution, consider $V^{\ast}(\bs{\theta}^{(t)}, \mathbf{\Phi}^{(t)}, \mathcal{T})$ which is the maximally achievable expected total future reward with current information $\bs{\theta}^{(t)}$, $\mathbf{\Phi}^{(t)}$ and remaining steps ($\mathcal{T} = 1,2 $). With the updated belief according to Eq. (\ref{eq:cond_ri}) already \textit{given}, the optimal expected reward for the second stage is simply a greedy approach:
\begin{align}\label{eq:V1}
V_{\text{CU}}^\ast(\bs{\theta}^{(2)}, \mathbf{\Phi}^{(2)}, 1)&= \max_{\pi}U_\pi^{(2)} \nonumber\\
&= \max_{\bs{v}\subset \mathcal{U}\bl \bs{u}}\mathbb{E}^{(2)}[\bs{1}^T\bs{R_v}]\nonumber\\
&=\max_{\bs{v} \subset \mathcal{U}\bl \bs{u}} \bs{1}^T \bs{\theta_v}^{(2)}.
\end{align}

By working backwards the total maximal expected reward for two stages can be obtained as
\begin{align}\nonumber
&V_{\text{CU}}^\ast(\bs{\theta}^{(1)},  \mathbf{\Phi}^{(1)}, 2) =\max_{\pi} (U_{\pi}^{(1)} +U_{\pi}^{(2)} )\\
=&\max_{\bs{u}\subset \mathcal{U}}\left(\mathbb{E}^{(1)}[\bs{1}^T\bs{R_u} + V_{\text{CU}}^\ast(\bs{\theta}^{(2)}, \mathbf{\Phi}^{(2)}, 1)] \right) \label{eq:mg_value_function}\\
= &\max_{\bs{u}\subset \mathcal{U}}\bigg(\mathbb{E}^{(1)}[\bs{1}^T\bs{R_u}] +
\int p^{(1)}(\bs{R_u}=\bs{r_u})V_{\text{CU}}^\ast(\bs{\theta}^{(2)}, \mathbf{\Phi}^{(2)}, 1)d \bs{r_u}\bigg). \nonumber
\end{align}

Substituting  Eqs. (\ref{eq:V1}) and (\ref{eq:update_mu}) into (\ref{eq:mg_value_function}) we reach the exact solution obtained by value iteration:
\begin{align}
\nonumber
&V_{\text{CU}}^\ast(\bs{\theta}^{(1)},\mathbf{\Phi}^{(1)}, 2) = \max_{\bs{u}\subset \mathcal{U}}\bigg\{ \overbrace{\bs{1}^T\bs{\theta_u}^{(1)}}^{\text{exploitation}} + \\
&\underbrace{\int p^{(1)}(\bs{R_u}=\bs{r_u})\max_{\bs{v}\subset \mathcal{U}\bl \bs{u}}\bigg[ \bs{1}^T \bigg(\bs{\theta}^{(1)}_{\bs{v}} + \mathbf{\Phi}^{(1)}_{\bs{v,u}}
[\mathbf{\Phi}^{(1)}_{\bs{u,u}}]^{-1}(\bs{r_u}-\bs{\theta}^{(1)}_{\bs{u}})\bigg)\bigg] d \bs{r_u}}_{\text{exploration}}\bigg\}.
\label{eq:mg_value_fun}
\end{align}

Eq. (\ref{eq:mg_value_fun}) suggests that the merit of choosing users $\bs{u}$ at the initial stage lies in two components:

\begin{itemize}
\item \textbf{Exploitation}. It is the immediate expected reward, denoted by $\bs{1}^T{\bs{\theta_u}^{(1)}}$, determined by the prior information on the users.

\item \textbf{Exploration}. The exploration component shows how the feedback from users $\bs{u}$ can lead the system to find optimal selections with updated knowledge. Consider that the feedback deviates from the prior information such that $(\bs{r_u}-\bs{\theta_u}^{(1)})\neq \bs{0}$, the updated belief state will then lead us to find users which bring ``extra'' returns via the term $\mathbf{\Phi}^{(1)}_{\bs{v,u}}
[\mathbf{\Phi}^{(1)}_{\bs{u,u}}]^{-1}(\bs{r_u}-\bs{\theta}^{(1)}_{\bs{u}})$. No matter the deviation is positive or negative, we can always benefit from it by selecting corresponding optimal users in the second stage.
As mentions above, this term relates to correlations between the
users of the two stages. The larger the correlations are, the more the system can \textit{gain} from the discrepancy between the observations and the prior information.
\end{itemize}

\vspace{0.1cm}

\subsection{Matrix Factorization Model with POMDP}
\label{sec:mf_pomdp}
\begin{figure}[t]
 \centering
   \includegraphics[width=3.5in]{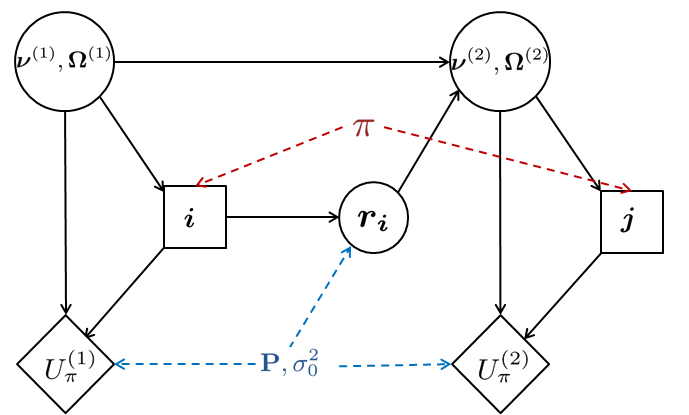}
   \captionsetup{width=0.85\textwidth}
 \caption{The two-stage MF-POMDP as illustrated by an influence diagram, with respect to
 the matrix factorization model. }
 \label{fig:mf_pomdp}
\end{figure}

To gain insights from the formulation of latent factor models,
consider MF with POMDP (MF-POMDP). For this purpose, we use the probabilistic model
$\bs{R} = \mathbf{P}\bs{q} + \xi$ such that
$\mathbf{P} = (\bs{p}_1, \bs{p}_2, \ldots,
\bs{p}_{|\mathcal{U}|})^T$
is a $|\mathcal{U}|\times K$ matrix containing the users' information, $\bs{q}$
is a $K$-dimensional item vector, and $\xi$ is a random variable
with zero mean and variance $\sigma_0^2$. If we assume fixed user
vectors $\mathbf{P}$ and unknown item vector $\bs{q}$ \cite{zhao2013interactive,shi2012adaptive}, CU-POMDP is
translated to a decision process under the belief state of the
unobservable item vector (see Figure~\ref{fig:mf_pomdp})
\begin{align}
p^{(t)}(\bs{q}) \sim \mathcal{N}(\bs{\nu}^{(t)},\mathbf{\Psi}^{(t)}),
\end{align}
where
$\bs{\nu}^{(1)}$ and $\mathbf{\Psi}^{(1)}$ are the mean and covariance matrix of the item vector. The belief state over the item vector then determines the belief over the preferences of users
\begin{align}
&p^{(t)}(\bs{R}) \sim \mathcal{N}(\mathbf{P}\bs{\nu}^{(t)}, \mathbf{P}\mathbf{\Psi}^{(t)}\mathbf{P}^T + \sigma_0^2 \mathbf{I}).
\label{eq:mf_r}
\end{align}

By observing users $\bs{u}$ with feedback $\bs{r_u}$ the belief state can be updated according to the Bayes rule
\begin{align}
&p^{(2)}(\bs{q}) \sim \mathcal{N}(\bs{\nu}^{(2)},\mathbf{\Psi}^{(2)}),\text{~where~} \\
&\bs{\nu}^{(2)}
= \bs{\nu}^{(1)}+\mathbf{\Psi}^{(1)}\mathbf{P}_{\bs{u}}^T(\mathbf{P}_{\bs{u}}\mathbf{\Psi}^{(1)}\mathbf{P}^T_{\bs{u}}+\sigma_0^2\mathbf{I})^{-1}(\bs{r_u}-\mathbf{P}_{\bs{u}}\bs{\nu}^{(1)}), \label{eq:update_nu} \\
&\bs{\Psi}^{(2)} =[(\mathbf{\Psi}^{(1)})^{-1}+\mathbf{P}^T_{\bs{u}}\mathbf{P}_{\bs{u}}/\sigma_0^2]^{-1}.
\end{align}

Thus,
\begin{align}\label{eq:update_mu_r}
&\mathbb{E}^{(2)}(\bs{R_{\bl u}}| \bs{r_u}) = \mathbf{P}_{\bs{\bl u}}\bs{\nu}^{(2)} \\
&= \mathbf{P}_{\bs{\bl u}}\bs{\nu}^{(1)}+\mathbf{P}_{\bs{\bl u}}\mathbf{\Psi}^{(1)}\mathbf{P}_{\bs{u}}^T(\mathbf{P}_{\bs{u}}\mathbf{\Psi}^{(1)}\mathbf{P}^T_{\bs{u}}+\sigma_0^2\mathbf{I})^{-1}(\bs{r_u}-\mathbf{P}_{\bs{u}}\bs{\nu}^{(1)}).
\nonumber
\end{align}

Comparing Eq. (\ref{eq:update_mu_r}) with Eq. (\ref{eq:update_mu}) we
find a nice alignment between the two models. Actually, by dimension
reduction the covariance between  user $u$'s and user $v$'s ratings can be translated as
\begin{align}
\label{eq:corr_bayes}
\Phi^{(1)}_{u,v} = \bs{p}_u^T \mathbf{\Psi}^{(1)} \bs{p}_v,
\end{align}
when $\sigma_0^2$ is very small compared to the covariance between the two users's true preferences ($\sigma_0^2 << \bs{p}_u^T \mathbf{\Psi}^{(1)} \bs{p}_v$). Eq. (\ref{eq:corr_bayes}) has converted the statistical property (the covariance of preferences between the two users) into the function of the feature vectors of the two users.

\vspace{-3pt}
By the same token, we write the optimal value function for the MF-POMDP as
\begin{align}
V_{\text{MF}}^\ast(&\bs{\nu}^{(1)},\mathbf{\Psi}^{(1)},2)=
\max_{\bs{u}\subset \mathcal{U}}\bigg\{ \bs{1}^T \mathbf{P}_{\bs{u}} \bs{\nu}^{(1)} + \nonumber\\
&\int p^{(1)}(\bs{R_u} = \bs{r_u}) \max_{\bs{v}\subset\mathcal{U}\bl\bs{u} }\bigg[\bs{1}^T\bigg( \mathbf{P}_{\bs{v}} \bs{\nu}^{(1)} + \label{eq:mf_value_function}\\ &\mathbf{P}_{\bs{v}}\mathbf{\Psi}^{(1)}\mathbf{P}_{\bs{u}}^T[\mathbf{P}_{\bs{u}}
\mathbf{\Psi}^{(1)}\mathbf{P}_{\bs{u}}^T + \sigma_0^2\mathbf{I}]^{-1}(\bs{r_u} - \mathbf{P}_{\bs{u}}\bs{\nu}^{(1)})\bigg)d\bs{r_u}
\bigg]\bigg\}.
\nonumber
\end{align}

\subsection{A Toy Example}

\vspace{-0.1cm}
Let us look at a simple three-user case and its analytical solution. In this example, one user is selected in each stage. We  base this example on the CU model so that the effect of user-user correlation can be illustrated more straightforwardly.

Suppose
\begin{footnotesize}
\begin{align}
\nonumber
\bs{\theta}^{(1)} =  \left(
\begin{array}{c}
\theta^{(1)}_1 \\
\theta^{(1)}_2 \\
\theta^{(1)}_3
\end{array}\right), ~~
\mathbf{\Phi}^{(1)} = \left(
\begin{array} {cccc}
\Phi^{(1)}_{1,1} & \Phi^{(1)}_{1,2} & \Phi^{(1)}_{1,3}  \\\nonumber
\Phi^{(1)}_{2,1} & \Phi^{(1)}_{2,2} & \Phi^{(1)}_{2,3}  \\\nonumber
\Phi^{(1)}_{3,1} & \Phi^{(1)}_{3,2} & \Phi^{(1)}_{3,3}  \nonumber
\end{array}\right).
\end{align}
\end{footnotesize}

Without loss of generality, we assume $\Phi^{(1)}_{1,3} > \Phi^{(1)}_{1,2} > \Phi^{(1)}_{2,3}$ (and ignore the case with equal covariance for now). Suppose user 1 is selected  in the initial stage with the observation as $r_1$, the update for the second and the third users are,
\begin{align}
\nonumber
\theta^{(2)}_2(r_1) &= \theta^{(1)}_2 + \Phi^{(1)}_{2,1}(\Phi^{(1)}_{1,1})^{-1}(r_1-\theta^{(1)}_1), \\ \nonumber
\theta^{(2)}_3(r_1) &= \theta^{(1)}_3 + \Phi^{(1)}_{3,1}({\Phi^{(1)}_{1,1}})^{-1}(r_1-\theta^{(1)}_1).
\end{align}

By introducing $z_1 = (r_1 - \theta^{(1)}_1)/\sqrt{\Phi^{(1)}_{1,1}}$, the above updates become
\begin{align}
\nonumber
\theta^{(2)}_2(z_1) &= \theta^{(1)}_2 + \Phi^{(1)}_{2,1}({\Phi^{(1)}_{1,1}})^{-1/2} z_1, \\ \nonumber
\theta^{(2)}_3(z_1) &= \theta^{(1)}_3 + \Phi^{(1)}_{3,1}({\Phi^{(1)}_{1,1}})^{-1/2}z_1.
\end{align}

We can see that both $\theta^{(2)}_2$ and $\theta^{(2)}_3$ are linear in $z_1$. The turning point between choosing user $2$ and user $3$ is obtained when the above two are equal to each other, which is at
\begin{align*}
d_{1} = \frac{\theta^{(1)}_2-\theta^{(1)}_3}{\Phi^{(1)}_{3,1}-\Phi^{(1)}_{2,1}} \sqrt{\Phi^{(1)}_{1,1}}.
\end{align*}

Since $\Phi^{(1)}_{3,1}>\Phi^{(1)}_{2,1}$, if $z_1 > d_1$, user 3 should be selected whereas if $z_1 < d_1$ user 2 should be selected in the second stage. Thus, the optimal reward when choosing user $1$ at the initial stage is

\begin{align*}
V^\ast_{u = 1}(&\theta^{(1)}, \Phi^{(1)}, 2) =\\
 \theta^{(1)}_1& + \int p^{(1)}(r_1)\cdot \max_{v = 2,3} \left(\theta^{(1)}_v + \Phi^{(1)}_{v,1}(\Phi^{(1)}_{1,1})^{-1}(r_1- \theta^{(1)}_1)\right) dr_1 \\ \nonumber
=& \theta^{(1)}_1 + \int_{-\infty}^{d_{1}}  p^{(1)}(z_1)\left[\theta^{(1)}_2 + \Phi^{(1)}_{2,1}(\Phi^{(1)}_{1,1})^{-1/2} z_1\right]dz_1  \\
&+\int_{d_{1}}^{\infty}p^{(1)}(z_1) \left[ \theta^{(1)}_3 + \Phi^{(1)}_{3,1}({\Phi^{(1)}_{1,1}})^{-1/2} z_1
 \right] dz_1 \\
=& \theta^{(1)}_1 + 1/2(\theta^{(1)}_2 + \theta^{(1)}_3) + 1/2(\theta^{(1)}_2-\theta^{(1)}_3)\textbf{erf}(\frac{d_1}{\sqrt{2}}) \\
&- \frac{1}{\sqrt{2\pi}}\frac{\Phi^{(1)}_{2,1}-\Phi^{(1)}_{3,1}}{\sqrt{\Phi^{(1)}_{1,1}}}\textbf{e}^{-\frac{d_1^2}{2}}.
\end{align*}

Similarly,
\begin{align*}
V^\ast_{u=2}(\theta^{(1)}, &\Phi^{(1)}, 2) = \\
 \theta^{(1)}_2& + \int p^{(1)}(r_2)\cdot \max_{v =3,1} \left(\theta^{(1)}_v + \Phi^{(1)}_{v,2}(\Phi^{(1)}_{2,2})^{-1}(r_2- \theta^{(1)}_2)\right) dr_2 \\ \nonumber
= &\theta^{(1)}_2 + \int_{-\infty}^{d_{2}}  p^{(1)}(z_2)\left[\theta^{(1)}_3 + \Phi^{(1)}_{3,2}(\Phi^{(1)}_{2,2})^{-1/2} z_2\right]dz_2  \\
&+\int_{d_{2}}^{\infty}p^{(1)}(z_2) \left[ \theta^{(1)}_1 + \Phi^{(1)}_{1,2}({\Phi^{(1)}_{2,2}})^{-1/2} z_2
 \right] dz_2\\
& =  \theta^{(1)}_2 + 1/2(\theta^{(1)}_3 + \theta^{(1)}_1) + 1/2(\theta^{(1)}_3-\theta^{(1)}_1)\textbf{erf}(\frac{d_2}{\sqrt{2}}) \\ &-\frac{1}{\sqrt{2\pi}}\frac{\Phi^{(1)}_{3,2}-\Phi^{(1)}_{1,2}}{\sqrt{\Phi^{(1)}_{2,2}}}\textbf{e}^{-\frac{d_2^2}{2}},\\
\end{align*}
\begin{align*}
V^\ast_{u=3}(\theta^{(1)}, &\Phi^{(1)}, 2)= \\
 \theta^{(1)}_3& + \int p^{(1)}(r_3)\cdot \max_{v=1,2} \left(\theta^{(1)}_v + \Phi^{(1)}_{v,3}(\Phi^{(1)}_{3,3})^{-1}(r_3- \theta^{(1)}_3)\right) dr_3 \\ \nonumber
=& \theta^{(1)}_3 + \int_{-\infty}^{d_{3}}  p^{(1)}(z_3)\left[\theta^{(1)}_2 + \Phi^{(1)}_{2,3}(\Phi^{(1)}_{3,3})^{-1/2} z_3\right]dz_3  \\
&+\int_{d_{3}}^{\infty}p^{(1)}(z_3) \left[ \theta^{(1)}_1 + \Phi^{(1)}_{1,3}({\Phi^{(1)}_{3,3}})^{-1/2} z_3
 \right] dz_3\\
&=  \theta^{(1)}_3 + 1/2(\theta^{(1)}_1 + \theta^{(1)}_2) +1/2(\theta^{(1)}_2-\theta^{(1)}_1)\textbf{erf}(\frac{d_3}{\sqrt{2}}) \\ &-\frac{1}{\sqrt{2\pi}}\frac{\Phi^{(1)}_{2,3}-\Phi^{(1)}_{1,3}}{\sqrt{\Phi^{(1)}_{3,3}}}\textbf{e}^{-\frac{d_3^2}{2}},\end{align*}
where
\begin{align*}
d_{2} = \frac{\theta^{(1)}_3-\theta^{(1)}_1}{\Phi^{(1)}_{1,2}-\Phi^{(1)}_{3,2}} \sqrt{\Phi^{(1)}_{2,2}}, ~d_{3} = \frac{\theta^{(1)}_2-\theta^{(1)}_1}{\Phi^{(1)}_{1,3}-\Phi^{(1)}_{2,3}} \sqrt{\Phi^{(1)}_{3,3}}.
\end{align*}

Note that the above formula are not rotational symmetric due to the asymmetry caused by $\Phi^{(1)}_{1,3} > \Phi^{(1)}_{1,2} > \Phi^{(1)}_{2,3}$.

\vspace{10pt}
To illustrate the results, let us look at a numerical example according to the above solutions. Suppose
\begin{footnotesize}
\begin{align*}
\nonumber
{\bs{\theta}}^{(1)}
= \left(
\begin{array}{c}
3.2 \\
2.5 \\
3.5
\end{array}
\right), ~~
{\mathbf{\Phi}}^{(1)}
=\left(
\begin{array} {ccc}
1.6 & 0.25 & 1.6  \\
0.25 & 3.2 & 0.20 \\
1.6 & 0.20 & 3.5
\end{array}
\right).
\end{align*}
\end{footnotesize}

The correlation matrix is thus
\begin{footnotesize}
\begin{align*}
\nonumber
{\mathbf{C}}^{(1)}
=\left(
\begin{array} {ccc}
1 & 0.11 & 0.68  \\
0.11 & 1 & 0.06 \\
0.68 & 0.06 & 1
\end{array}
\right).
\end{align*}
\end{footnotesize}

When user 1 is selected at the initial stage:
\begin{align*}
\nonumber
\theta^{(2)}_2 (r_1)&= \theta^{(1)}_2 + \Phi^{(1)}_{2,1}(\Phi^{(1)}_{1,1})^{-1}(r_1-\theta^{(1)}_1) \\
& = 2.5+0.25 \times (1.6)^{-1}(r_1-3.2),\\ \nonumber
\theta^{(2)}_3(r_1) &= \theta^{(1)}_3 + \Phi^{(1)}_{3,1}({\Phi^{(1)}_{1,1}})^{-1}(r_1-\theta^{(1)}_1) \\
&= 3.5+1.6 \times(1.6)^{-1}(r_1-3.2).
\end{align*}

Therefore, when $r_1< 2.01$ we should choose user 2 in the second stage whilst when $r_1> 2.01$ we should choose user 3 (when $r_1 = 2.01$ choosing either will give the same expected reward in the second stage).
The corresponding value function is

\begin{align*}
\nonumber
&V^\ast_{u=1}(\bs{\theta}^{(1)}, \mathbf{\Phi}^{(1)}, 2) \\
&= \theta^{(1)}_1 + \int p^{(1)}(r_1)\cdot \max_{j =2,3} \left(\theta^{(1)}_v + \Phi^{(1)}_{j,1}(\Phi^{(1)}_{1,1})^{-1}(r_1- \theta^{(1)}_1)\right) dr_1 \\
&=  3.2 + \int_{- \infty}^{2.01} p^{(1)}(r_1)(2.5+0.25 \times (1.6)^{-1}(r_1-3.2)) dr_1 \\
&+\int_{2.01}^{+\infty}p^{(1)}(r_1)(3.5+1.60 \times(1.6)^{-1}(r_1-3.2))dr_1 \\
&\approx 6.80.
\end{align*}

Similarly, we can obtain the value functions for choosing user 2 and 3 at the initial stage
\begin{align*}
V^\ast_{u=2}(\bs{\theta}^{(1)}, \mathbf{\Phi}^{(1)}, 2) &\approx 5.7,\\
V^\ast_{u=3}(\bs{\theta}^{(1)}, \mathbf{\Phi}^{(1)}, 2) &\approx 6.77.
\end{align*}

And thus obtain the final value function
\begin{align*}
V^\ast(\bs{\theta}^{(1)}, \mathbf{\Phi}^{(1)}, 2) = \max(6.80, 5.7,6.77) = 6.80.
\end{align*}

We can see that the value function favours the first user at the first step, even
though the prior information about the users favours the third user over the
first user. Due to the fact that user $1$ is highly correlated to user 3, and is more correlated with
user 2 than user 3 is, choosing user $1$ at the initial stage will enable the system to judge better in the second stage which results in a higher total expected reward over the two stages.

\subsection{Computational Complexity}
The exact solution of a finite-horizon POMDP has been proven to be PSPACE-complete \cite{papadimitriou1987complexity}. In our case, the decision space at the initial stage is $C^{|\mathcal{U}|}_{m}$.
For each decision, the $m$-dimensional observation space will be divided into $C^{|\mathcal{U}|-m}_{n}$ regions, each region corresponds to a (possibly) different optimal user combination to choose for the second stage. That is, the exact solution suggested by the value iteration algorithm requires going through all the possible decisions and all possible observations, which is intractable.

\section{Approximation}
\label{sec:approximation}
To ease the intractability of the exact solution, we propose an approximation solution here, named guided exploitation-exploration (GEE).
We provide its form for both the CU model and the MF model below.

\subsection{Approximation for CU-POMDP}
\label{sec:mg_gee}
From Section \ref{sec:mg_value_iteration}, we have seen that the merit of selecting a group of users lies
both in the immediate reward term (the exploitation part of Eq. (\ref{eq:mg_value_fun})) and in how it can guide the system to find promising users in the next stage through the system update (the exploration part of Eq. (\ref{eq:mg_value_fun})).
However,  when the decision of the initial stage is made, the system's belief state update is unknown before receiving any observations. To investigate the influence of selecting users $\bs{u}$ only (before making any observations),  let us consider the conditional distribution of unselected users $\bl \bs{u}$ over the selection of users $\bs{u}$, $p(\bs{R}_{\bl \bs{u}}|\bs{u})$. Note that this conditional distribution is different from Eq. (\ref{eq:cond_ri}) because
it is the distribution conditioned on the action $\bs{u}$ instead of the observations, as at the initial-decision stage these observations are still unknown.

\vspace{10pt}
Because the observations are not made yet, the expected feedback conditioned on the selection remains unchanged
\begin{align}
&\mathbb{E}[\bs{R_{\bl u}}|\bs{u}] = \bs{\theta_{\bl u}}^{(1)}.
\end{align}

\vspace{10pt}
However, its covariance changes according to the choice of $\bs{u}$:
\begin{align}
\cov[\bs{R_{\bl u}}|\bs{u}] &= \cov\left[ \bs{\theta_u}^{(1)} + \mathbf{\Phi}^{(1)}_{\bl \bs{u},\bs{u}}(\mathbf{\Phi}^{(1)}_{\bs{u,u}})^{-1}
(\bs{R}_{\bs{u}}-\bs{\theta}^{(1)}_{\bs{u}}) \right]\nonumber\\
& =  \mathbf{\Phi}^{(1)}_{{\bl \bs{u},\bs{u}}}(\mathbf{\Phi}^{(1)}_{\bs{u,u}})^{-1} \cov(\bs{R_u}) (\mathbf{\Phi}^{(1)}_{\bs{u,u}})^{-1}\mathbf{\Phi}^{(1)}_{{\bs{u},\bl \bs{u}}}\nonumber \\
& =  \mathbf{\Phi}^{(1)}_{{\bl \bs{u},\bs{u}}}(\mathbf{\Phi}^{(1)}_{\bs{u,u}})^{-1} \mathbf{\Phi}^{(1)}_{{\bs{u},\bl \bs{u}}},
\label{eq:convar}
\end{align}
where the last step is due to $\cov(\bs{R_u}) = \mathbf{\Phi}_{\bs{u,u}}^{(1)}$.

\vspace{10pt}
Therefore, with the initial-stage users as $\bs{u}$, the expected returns at the second stage by choosing users $\bs{v}$ are bounded by the interval $\Theta_{\bs{u},\bs{v}}$:
\begin{align}
\nonumber
\Theta_{\bs{u},\bs{v}} = &\bigg[\bs{1}^T\left( \bs{\theta_v}^{(1)}-\lambda  \cdot \text{diag}\bigg[\text{Dg}^{-\frac{1}{2}} \left(\cov (\bs{R}_{\bs{v}}|\bs{u})\right)\bigg]\right),\\
~~~~~~~&\bs{1}^T\left(\bs{\theta}^{(1)}_{\bs{v}}+\lambda \cdot \text{diag}\bigg[\text{Dg}^{-\frac{1}{2}}  \left(\cov (\bs{R}_{\bs{v}}|\bs{u})\right)\bigg] \right)\bigg]
\end{align}
with the probability at least $(1-2 \mathbf{e}^{-\lambda^2/2})^n$ \cite{viens2009stein}\footnote{To be more exact, the conditional vector $\bs{R_{\bl u}}|\bs{u}$ is bounded in an ellipsoid. This form is obtained with an approximation of considering only the diagonal elements of $\cov(\bs{R}_{\bl \bs{u}}|\bs{u})$.}.

The GEE algorithm therefore optimistically assumes the highest return could be achieved within this interval \cite{walsh2009exploring}.
And thus we choose the users $\bs{u}$ which can achieve the highest total ratings under this assumption
\begin{align}
\nonumber
\pi&_{\text{CU-GEE}}(\bs{\theta}^{(1)},\mathbf{\Phi}^{(1)},\mathcal{U})\\
=&\argmax_{\bs{u}\subset \mathcal{U}}\bigg\{\bs{1}^T\bs{\theta_u}^{(1)} +
\max_{\bs{v} \subset \mathcal{U} \bl \bs{u} }\bs{1}^T\bigg(\bs{\theta}^{(1)}_{\bs{v}}+ \nonumber \\
&\lambda \cdot \text{diag}\bigg[\text{Dg}^{-\frac{1}{2}}  \left( \mathbf{\Phi}^{(1)}_{{\bs{v},\bs{u}}}
(\mathbf{\Phi}^{(1)}_{\bs{u,u}})^{-1} \mathbf{\Phi}^{(1)}_{{\bs{u},\bs{v}}}\right)\bigg] \bigg)\bigg\}.
\label{eq:mg_gee}
\end{align}

\begin{algorithm}[t]
\caption{CU-GEE by Sampling}
\begin{algorithmic}
\REQUIRE{Prior mean ratings $\bs{\theta}^{(1)}$, covariance matrix $\bs{\Phi}^{(1)}$, GEE parameter $\lambda$, available users $\mathcal{U}$}
\STATE Initialise $\bs{u}^* \leftarrow \emptyset$\\
\FOR {$t=1 \ldots T$}
        \STATE Sample $\bs{u}_t$ ($|\bs{u}_t|=m$) from $\mathcal{U}$
        \STATE Calculate
        $V^{\text{CU-GEE}}_{\bs{u}_t}$
        according to Eq. (\ref{eq:mg_gee})
\IF {$V_{\bs{u}_t}^{\text{CU-GEE}}$ is the largest so far}
            \STATE Update $\bs{u}^* \leftarrow \bs{u}_t$
        \ENDIF
\ENDFOR
\end{algorithmic}
\label{alg:mg_gee}
\end{algorithm}

This algorithm suggests that, in order to determine the users for stage one, we first calculate the immediate reward based on the prior information. Then we calculate the optimistic reward when acting optimally in the second stage. We call GEE \textit{guided} as the initial-stage decision is optimistically guided by pseudo optimal user selections in the next stage. By inspecting into the next stage, we utilise the correlation between users of the two stages, which will be explained further in Section \ref{sec:mg_gee_I}. To implement this algorithm, we can adopt a sampling-based method depicted in Algorithm \ref{alg:mg_gee}.

\subsubsection{\textbf{Independent Intra-Stage User Assumption}}
\label{sec:mg_gee_I}

To align our algorithm with the popular memory-based CF, we adopt the
correlation function Eq. (\ref{eq:corr}) and reformulate Eq. (\ref{eq:mg_gee}) as follows:
\begin{align}
&\mathbf{\Phi}^{(1)}_{{\bs{v},\bs{u}}}
(\mathbf{\Phi}^{(1)}_{\bs{u,u}})^{-1} \mathbf{\Phi}^{(1)}_{{\bs{u},\bs{v}}} \nonumber \\
 =
&\big[\text{diag}(\bs{\phi_v}^{(1)}) \mathbf{C}_{\bs{v,u}}^{(1)} \text{diag}(\bs{\phi_u}^{(1)})\big]\big[ \text{diag}^{-1}(\bs{\phi_u}^{(1)})   (\mathbf{C}_{\bs{u,u}}^{(1)})^{-1}  \text{diag}^{-1}(\bs{\phi_u}^{(1)})\big]  \big[\text{diag}(\bs{\phi_u}^{(1)})  \mathbf{C}_{\bs{u,v}}^{(1)} \text{diag}(\bs{\phi_v}^{(1)}) \big]\nonumber \\
=& \text{diag}(\bs{\phi}_{\bs{v}}^{(1)}) \mathbf{C}_{\bs{v,u}}^{(1)} (\mathbf{C}_{\bs{u,u}}^{(1)})^{-1}  \mathbf{C}_{\bs{u,v}}^{(1)}   \text{diag}(\bs{\phi_v}^{(1)}).
\end{align}

Eq. (\ref{eq:mg_gee})  thus becomes
\begin{align}
\nonumber
&\pi_{\text{CU-GEE}'}(\bs{\theta}^{(1)}, \bs{\phi}^{(1)},\mathbf{C}^{(1)}, \mathcal{U})\\
& = \argmax_{\bs{u}\subset \mathcal{U}}\bigg\{ \bs{1}^T \bs{\theta_u}^{(1)} + \max_{\bs{v} \subset \mathcal{U} \bl \bs{u}} \bs{1}^T \bigg(\bs{\theta_v}^{(1)}+ \nonumber\\
&\lambda \cdot \text{diag}\bigg[\text{Dg}^{-\frac{1}{2}} \left( \text{diag}(\bs{\phi_v}^{(1)})
\mathbf{C}^{(1)}_{\bs{v},\bs{u}} (\mathbf{C}^{(1)}_{\bs{u},\bs{u}})^{-1} \mathbf{C}^{(1)}_{\bs{u},\bs{v}}\text{diag}(\bs{\phi_v}^{(1)})
 \right)\bigg]\bigg\}. \label{eq:mg_gee_corr}
\end{align}

\vspace{10pt}
The term of $(\mathbf{C}^{(1)}_{\bs{u},\bs{u}})^{-1}$ in the above equation suggests us to diversify the items in the initial stage. Here in order to catch the more important relation between the two stages, we assume the initial-stage users $\bs{u}$ are independent of each other, which suggests an already-diversified user list.
In addition to the independent assumption, we also impose an equal variance assumption, i.e., all the users have the same variance $\phi'^2$ (so $\text{diag}(\bs{\phi_v}^{(1)}) = \phi'\mathbf{I}$). With the two assumptions, Eq. (\ref{eq:mg_gee_corr}) can be further approximated to
 \begin{align}
\nonumber
 &\pi_{\text{CU-GEE-I}}(\bs{\theta}^{(1)},\mathbf{C}^{(1)},\mathcal{U}) \\
   =& \argmax_{\bs{u}\subset \mathcal{U}}\left[ \sum_{\alpha =1}^{m}\theta^{(1)}_{u_{\alpha}} + \max_{\bs{v} \subset \mathcal{U}\bl \bs{u} }\sum_{\beta = 1}^{n}  \left( \theta^{(1)}_{v_{\beta}}+\lambda'\sqrt{\sum_{\alpha = 1}^{m}(\rho^{(1)}_{u_{\alpha},v_{\beta}})^2 }\right) \right],
 \label{eq:mg_gee_I}
 \end{align}
 where $\lambda' = \lambda\phi'$, and $\rho^{(1)}_{u_{\alpha},v_{\beta}}$ is just the correlation between $u_{\alpha}$ and $v_{\beta}$ according to the prior information.
 The effect of inter-stage user-user correlations is shown clearly in the above formula. According to Eq. (\ref{eq:mg_gee_I}), given the user selection at the initial stage $\bs{u}$, we can foresee the optimistic return in the next stage through highly expected values (via $\theta_{v_{\beta}}^{(1)}$) and also highly correlated  users (via the term $\sqrt{\sum_{\alpha = 1}^{m}(\rho^{(1)}_{u_{\alpha},v_{\beta}})^2 }$). Identifying these users then guides the system to determine the user selection $\bs{u}^\ast$.

\begin{algorithm}[t]
\caption{CU-GEE-I by Sampling}
\begin{algorithmic}
\REQUIRE{Prior mean ratings $\bs{\theta}^{(1)}$, correlation matrix $\mathbf{C}^{(1)}$, GEE parameter $\lambda'$, available users $\mathcal{U}$}
\STATE Initialise $\bs{u}^* \leftarrow \emptyset$\\
\FOR {$t=1 \ldots T$}
        \STATE Sample $\bs{u}_t$ ($|\bs{u}_t|=m$) from $\mathcal{U}$
        \STATE Calculate $V_{\bs{u}_t}^{\text{CU-GEE-I}}$ according to Eq. (\ref{eq:mg_gee_I})
        \IF {$V_{\bs{u}_t}^{\text{CU-GEE-I}}$ is the largest so far}
            \STATE Update $\bs{u}^* \leftarrow \bs{u}_t$
        \ENDIF
\ENDFOR
\end{algorithmic}
\label{alg:mg_gee_I}
\end{algorithm}

\vspace{10pt}
The sampling method for this algorithm is illustrated in Algorithm \ref{alg:mg_gee_I}. 

\subsection{Approximation for MF-POMDP} 
\label{sec:mf_gee}
With the MF model, the conditional covariance matrix of $\bs{R_{\bl u}}$ given
the user selection $\bs{u}$ is written as
\begin{align}
\cov(\bs{R_{\bl u}}|\bs{u})&= \mathbf{P}_{\bl \bs{u}} \mathbf{\Psi}^{(1)}\mathbf{P}_{\bs{u}}^T(\mathbf{P}_{\bs{u}}\mathbf{\Psi}^{(1)}
\mathbf{P}^T_{\bs{u}}+\sigma_0^2\mathbf{I})^{-1}
\mathbf{P}_{\bs{u}}\mathbf{\Psi}^{(1)}\mathbf{P}_{\bl \bs{u}}^T.
\end{align}

Following the same reasoning as in Section \ref{sec:mg_gee}, we give the formulation for the matrix factorization model
\begin{align}
&\pi_{\text{MF-GEE}}(\bs{\nu}^{(1)},\mathbf{\Psi}^{(1)},\mathcal{U}) \nonumber \\
=& \argmax_{\bs{u}\subset \mathcal{U}} \bigg\{ \bs{1}^T\mathbf{P}_{\bs{u}}\bs{\nu}^{(1)} +
\max_{\bs{v} \subset \mathcal{U} \bl \bs{u}}\bs{1}^T  \bigg(  \mathbf{P}_{\bs{v}}\bs{\nu}^{(1)}+ \lambda \cdot \label{eq:mf_gee}\\
\text{diag}&\bigg[\text{Dg}^{-\frac{1}{2}} \left(\mathbf{P}_{\bs{v}} \mathbf{\Psi}^{(1)}\mathbf{P}_{\bs{u}}^T(\mathbf{P}_{\bs{u}}\mathbf{\Psi}^{(1)}
\mathbf{P}^T_{\bs{u}}+\sigma_0^2\mathbf{I})^{-1}
\mathbf{P}_{\bs{u}}\mathbf{\Psi}^{(1)}\mathbf{P}_{\bs{v}}^T \right)\bigg] \bigg)\bigg\} \nonumber
\end{align}

The corresponding algorithm is shown in Algorithm \ref{alg:mf_gee}.

\begin{algorithm}[t]
\caption{MF-GEE by Sampling}
\begin{algorithmic}
\REQUIRE{Prior mean $\bs{\nu}^{(1)}$ and covariance matrix $\mathbf{\Psi}^{(1)}$ of the target item feature vector, GEE parameter $\lambda$, available users $\mathcal{U}$}
\STATE Initialise $\bs{u}^* \leftarrow \emptyset$\\
\FOR {$t=1 \ldots T$}
        \STATE Sample $\bs{u}_t$ ($|\bs{u}_t|=m$) from $\mathcal{U}$
        \STATE Calculate $V_{\bs{u}_t}^{\text{MF-GEE}}$ according to Eq. (\ref{eq:mf_gee})
        \IF {$V_{\bs{u}_t}^{\text{MF-GEE}}$ is the largest so far}
            \STATE Update $\bs{u}^* \leftarrow \bs{u}_t$
        \ENDIF
\ENDFOR
\end{algorithmic}
\label{alg:mf_gee}
\end{algorithm}

\subsubsection{\textbf{Independent Intra-Stage User Assumption}}

With the MF model, in addition to the independent intra-stage user assumption which turns $\mathbf{P}_{\bs{u}}\mathbf{\Psi}^{(1)}\mathbf{P}_{\bs{u}}^T$ into a diagonal matrix,
we may also assume independent latent dimensions such that the prior covariance matrix is diagonal: $\mathbf{\Psi}^{(1)} = \text{diag}^2[\bs{\psi}^{(1)}]$, where $\bs{\psi}^{(1)}$ are the standard deviations of latent dimensions. Eq. (\ref{eq:mf_gee}) can be further
simplified as:
 \begin{align}
 \label{eq:mf_gee_1}
&\pi_\text{MF-GEE-I}(\bs{\nu}^{(1)}, \bs{\psi}^{(1)},\mathcal{U}) =\argmax_{\bs{u}\subset \mathcal{U}}\bigg\{
 \sum_{\alpha = 1}^m\bs{p}_{u_{\alpha}}^T\bs{\nu}^{(1)} +\nonumber\\
&\max_{\bs{v}\subset \mathcal{U} \bl \bs{u}}\sum_{\beta = 1}^n
\bigg(\bs{p}_{v_{\beta}}^T\bs{ \nu}^{(1)}+ \lambda\sqrt{ \sum_{\alpha = 1}^m \frac{(\bs{p}_{v_{\beta}}^T\text{diag}^2[\bs{\psi}^{(1)}] \bs{p}_{u_{\alpha}})^2}{\bs{p}^T_{u_{\alpha}}\text{diag}^2[\bs{\psi}^{(1)}]\bs{p}_{u_{\alpha}}+\sigma_0^2}}\bigg)\bigg\}.
 \end{align}

The corresponding algorithm is shown in Algorithm \ref{alg:mf_gee_I}.

\vspace{10pt}

Particularly, when assuming $\bs{\psi}^{(1)} = \psi^{(1)} \bs{1}$, i.e., equal
prior standard deviation (variance) along different dimensions, we gain the form
\begin{align}\nonumber
&\pi_\text{MF-GEE-II}(\bs{\nu}^{(1)}, \psi^{(1)},\mathcal{U}) =\argmax_{\bs{u}\subset \mathcal{U}}\bigg\{
 \sum_{\alpha = 1}^m\bs{p}_{u_{\alpha}}^T\bs{\nu}^{(1)} +\\
&\max_{\bs{v}\subset \mathcal{U} \bl  \bs{u}}\sum_{\beta = 1}^n
\bigg(\bs{p}_{v_{\beta}}^T\bs{ \nu}^{(1)}+ \lambda\sqrt{ \sum_{\alpha = 1}^m \frac{((\psi^{(1)})^2\bs{p}_{v_{\beta}}^T \bs{p}_{u_{\alpha}})^2}{(\psi^{(1)})^2 \bs{p}^T_{u_{\alpha}} \bs{p}_{u_{\alpha}}+\sigma_0^2}}\bigg)\bigg\}.
\label{eq:mf_gee_II}
\end{align}

Actually, with such a spherical prior variance, Eq. (\ref{eq:corr_bayes}) becomes $
\Phi^{(1)}_{u,v} =(\psi^{(1)})^2 \bs{p}_u^T \bs{p}_v$,
i.e., the covariance between $u$ and $v$ is proportional to the inner product of the user latent factors.
Actually, with a spherical prior variance, the correlation between user $u$ and $v$, $\rho_{u,v}$, is proportional to $\bs{p}_u^T \bs{p}_v$, corresponding to the MF obtained
 by a regularised linear regression estimation \cite{rubens2011active}.

\begin{algorithm}[t]
\caption{MF-GEE-I by Sampling}
\begin{algorithmic}
\REQUIRE{Prior mean $\bs{\nu}^{(1)}$ and the diagonal element of the covariance matrix $\bs{\psi}^{(1)}$ of the target item feature vector, GEE parameter $\lambda$, available users $\mathcal{U}$}
\STATE Initialise $\bs{u}^* \leftarrow \emptyset$\\
\FOR {$t=1 \ldots T$}
        \STATE Sample $\bs{u}_t$ ($|\bs{u}_t|=m$) from $\mathcal{U}$
        \STATE Calculate $V_{\bs{u}_t}^{\text{MF-GEE-I}}$ according to Eq. (\ref{eq:mf_gee_1})
        \IF {$V_{\bs{u}_t}^{\text{MF-GEE-I}}$ is the largest so far}
            \STATE Update $\bs{u}^* \leftarrow \bs{u}_t$
        \ENDIF
\ENDFOR
\end{algorithmic}
\label{alg:mf_gee_I}
\end{algorithm}

\section{Related Work and Discussion}
\label{sec:related-work}

\subsection{Collaborative Filtering}
Our work can be considered part of CF research
\cite{su2009survey}. CF provides efficient and personalised recommendations based on the
similarities between users and items. This can be achieved by mainly three approaches \cite{su2009survey}: a similarity-based approaches such as
neighbourhood based CF (user-based and item-based) \cite{herlocker1999algorithmic,deshpande2004item}, latent factor models  \cite{koren2009matrix,koren2008factorization,koren2008factorization,blei2003latent}, and hybrid methods \cite{burke2002hybrid}. We relate our work with the neighbourhood-based CF and latent factor models as follows.
\vspace{-7pt}

\subsubsection{Neighbourhood-Based CF}

Neighbourhood based CF provides a straightforward estimation of the target rating as a weighted summation of similar ratings: either the ratings from similar users, or the ratings to similar items, and can therefore provide explainable recommendations \cite{herlocker1999algorithmic,herlocker2002empirical,RecSystemSurvey,sarwar2001item,bell2007improved,rubens2011active}.

\vspace{10pt}
According to our analysis in Section \ref{sec:mg_pomdp}, neighbourhood-based models can be viewed as an approximate multivariate Gaussian preference model with the following two assumptions: (i) only the correlations between the target user and its neighbours are considered, and the neighbour users are assumed to be independent to each other; and (ii) all users have the same variance in their rating behaviours. In some practices,
the rating scores are also normalised with their standard deviations \cite{herlocker2002empirical}, which is referred to as the $Z$-score normalisation. We thus see the $Z$-score normalisation as a way to alleviate the prediction discrepancy caused by the second assumption. In addition, in practice, only the most-similar users are selected as neighbours, including top-$N$ filtering and threshold filtering strategies, to ease the computational cost \cite{rubens2011active}. Beside the Pearson correlation, Cosine vector similarity is also used, but it is argued that its performances are not as good as the Pearson correlation similarity measure \cite{breese1998empirical}.

\vspace{-2pt}
\subsubsection{Latent Factor Models}

Latent factor models
first project the user and item onto a latent feature space, and then base the score on the feature vectors of them. The correlations between user pairs are therefore translated as the vector similarity in the latent space (as shown in Section \ref{sec:mf_pomdp}).
Matrix factorisation is probably the most well-known method of latent factor  models \cite{koren2009matrix}. Singular value decomposition (SVD) \cite{koren2009matrix},
SVD++ \cite{koren2008factorization}, pLSA \cite{hofmann2004latent} and Latent Dirichlet Allocation \cite{blei2003latent} are among the more famous ones. Probabilistic matrix factorisation (MF) is one of the latent factors which is adopted in this paper \cite{salakhutdinov2008probabilistic} with respect its probabilistic property.

\subsection{Cold-Start Problems in CF}

Cold-start problems \cite{schein2002methods}
remain a major challenge for CF-based recommender systems, as
the prediction of ratings purely depends on the previously expressed user-item preferences without the use of any
content information, and for a new user or item this information is unavailable.

\vspace{10pt}
There is comparatively more literature on the user cold-start problems than the
item cold-start problems. For the former, an pre-recommendation `interview' process is usually adopted.
In an interview, the user first gives feedback on some questions provided by the system, such as preferences on some
popular items or highly informative items, or on a diversified list for the users to rate \cite{rashid2002getting,rashid2008learning}.
The interview phase can also be more intelligent, as decision-tree based methods suggest \cite{rashid2002getting,zhou2011functional,golbandi2011adaptive}.
AL forms an important branch for designing interview questions, and is most relevant to our approach. We have a thorough comparison in Section \ref{sec:active}.

\vspace{10pt}
In addition, many techniques
\cite{zhou2011functional,golbandi2011adaptive,zhao2013interactive}
assume an interactive process for sequential query selection,
i.e., only one query is chosen at a time for one user.  Then after the
response is collected, another query will be chosen according to the
user response to the previous query. In our case, as well as in many
other practical situations, multiple items or users should be
recommended in a batch manner to improve efficiency. Therefore,
iterative techniques are not applicable.

\subsection{Probabilistic Ranking Principle in CF}
The well-known probabilistic ranking principle (PRP) has been related
to CF \cite{wang2008probabilistic}, which
suggests that the top-$N$ recommendation list can be generated by
ranking according to the probability of relevance to the target (a
user or an item). Originated from information retrieval
\cite{robertson1977probability}, PRP implies documents to be ranked
in descending order by their probability of relevance can produce optimal
performance under the ``independent document'' assumption
\cite{Rijsbergen:1979:IR:539927}. In the item cold-start problem
scenario, supposing the rating is proportional to the relevance
probability, the list of users to recommend the item to should
be ranked according to the prior information of the users, such as the
rank of the user average ratings. On the other hand, in a user
cold-start scenario, the rank of recommended items should be the prior
average rating information of the items.

\vspace{10pt}
We have shown in this paper that PRP is not optimal as the correlations between users play an important role for the system to update, when considered as an interactive process. There are both intra-list correlations between users chosen in the initial stage and inter-list correlations between users in the first and remaining users (Eq. \ref{eq:mg_value_fun}). Especially, the inter-list correlations enable the system to update, and finally lead to more accurate predictions.

\subsection{Comparisons to Other EE Methods}

\subsubsection{Comparison to Active Learning}
\label{sec:active}
Active learning (AL) methods have been adopted to handle cold-start problems in recommender systems \cite{rubens2011active,harpale2008personalized,rubens2009output,rubens2007influence}, which are also referred to as optimal design by statisticians \cite{taguchi1986introduction}. AL uses a limited number of items (usually much smaller than the total number of available items) to present to the target user to review, and then learns the user's profile based on the users' feedback on these items. The criterion for selection is usually represented by a statistical measure such as achieving minimal mean squared error in the model estimation (A-optimality criterion) \cite{anava2015budget}, minimal 2-norm of the inverse of the information matrix (E-optimality criterion) \cite{rubens2009output} or minimal determinant of resulting covariance matrix of the system (D-optimality criterion) \cite{rubens2009output}. This objective differs from our objective function, and thus leads to significant differences from our approach.

\vspace{10pt}

There are two main differences between AL and our GEE approach.
First, AL techniques such as
D-Optimal design \cite{rubens2009output},
A-Optimal design \cite{anava2015budget} and their applications to the cold-start item problem
have divided  exploration  and  exploitation
 into two separate stages. In the exploration stage, a small number
of training points are selected for the system to learn, and in the exploitation stage
the gained information is fully exploited. However, the returns (or regrets) collected from
the exploration stage are not considered. In other words, The objective is imposed onto
only the exploitation stage, and thus the trade-off between exploration and
exploitation is not modeled \cite{rubens2011active}. For
example, in \cite{anava2015budget}, a budget has been imposed on the
number of users to select at the experimental stage, and these users' returns are
excluded from the objective function.

\vspace{10pt}
Second, the goal of AL is usually measured statistically using a global criterion. The criterion can be, for example, (to minimise) the mean square error of the estimates \cite{anava2015budget}, or, (to maximise) the differential Shannon information \cite{rubens2009output}. However, from Eqs. (\ref{eq:mg_value_fun}) and (\ref{eq:mf_value_function}) and from the example, we can see that the exact solution is achieved by prioritising the learning process towards the promising users of the next stage. Therefore, it is not necessary to achieve a global optimum. On the contrary, GEE captures this feature and make decisions guided by potential users of the second stage.

\subsubsection{Comparison to UCB methods}

The EE problem has been intensively studied in the literature of multi-armed bandit problems, where an agent decides dynamically which arm
 to choose at each step bearing the objective to maximise the total reward collected during a period of time \cite{auer2003using}. Gittins has provided an optimal solution under the condition that only one arm at a time can evolve  \cite{gittins2011multi}, but this is intractable in practice. UCB seeks a bounded regret instead of optimality and is  used to balance the exploitation and exploration in practice \cite{auer2003using,auer2002finite,srinivas2009gaussian,walsh2009exploring}. In UCB, usually a decision is made based on both the expectation and uncertainty of the return of individual choices at each step. In \cite{zhao2013interactive}, we proposed several UCB-based algorithms for a multiple-stage interactive recommendation process. And recently GP-UCB algorithms have also been applied to solve the user cold-start problems interactively in recommender systems \cite{vanchinathan2014explore}.

\vspace{10pt}
Our approach differs from UCB approaches in the following ways. First, UCB-based approaches seek to limit the regret within a bound, but they do not model how the specific selection within the bound can influence the outcome. In other words, EE achieved by UCB is not guided by the potential rewarding choice of the following stage, but is rather to limit the regret of the current stage. Second, UCB-based approaches are usually achieved in a long-term and interactive process, and may not be suitable for the two-stage process. Conversely, our algorithms are derived directly from the exact solutions of POMDP. They have directly considered the effect that choosing the initial-stage users has on the potential returns from the second stage.

\section{Experiment}
\label{sec:exp-twostage}
In this section, we compare our proposed approximate solutions with
several baseline methods. To understand the model further and
verify our theoretical analysis, we first present the results  on  synthetic data, and then on a real dataset.

\vspace{-5pt}

\subsection{Synthetic Data Experiment}
\label{sec:synthetic}

\subsubsection{\textbf{Synthetic Data Generation}}

 First, we define a $5$-dimensional latent space and randomly
generate a multivariate Gaussian distribution as the prior information of the cold-start item. In detail, each dimension of the multivariate Gaussian mean vector
is generated randomly according to $\mathcal{N}(0,0.1)$, and each dimension's standard deviation is generated according to $\mathcal{N}(0,1)$. Then we generate $50$ cold-start items according to this randomly-generated distribution.
Second, we generate 100 users' vectors according to $\mathcal{N}(\bs{0},\mathbf{I})$ as the available user pool for the $50$ cold-start items to target to. Their real ratings are then produced according to Eq. (\ref{eq:mf_r}) with the noise's standard deviation as $0.5$.
As such, we can obtain a $100\times 50$ rating matrix as the groundtruth. The true prior information is then provided for each compared algorithm to perform recommendations. Finally,
the above process is repeated for a total of $30$ times, each time with a different prior information of the cold-start items. The results are then  averaged over the different trials.

\begin{figure}[t]
\centering
	\subfigure[$N =10$]{
	\includegraphics[width=1.8in]{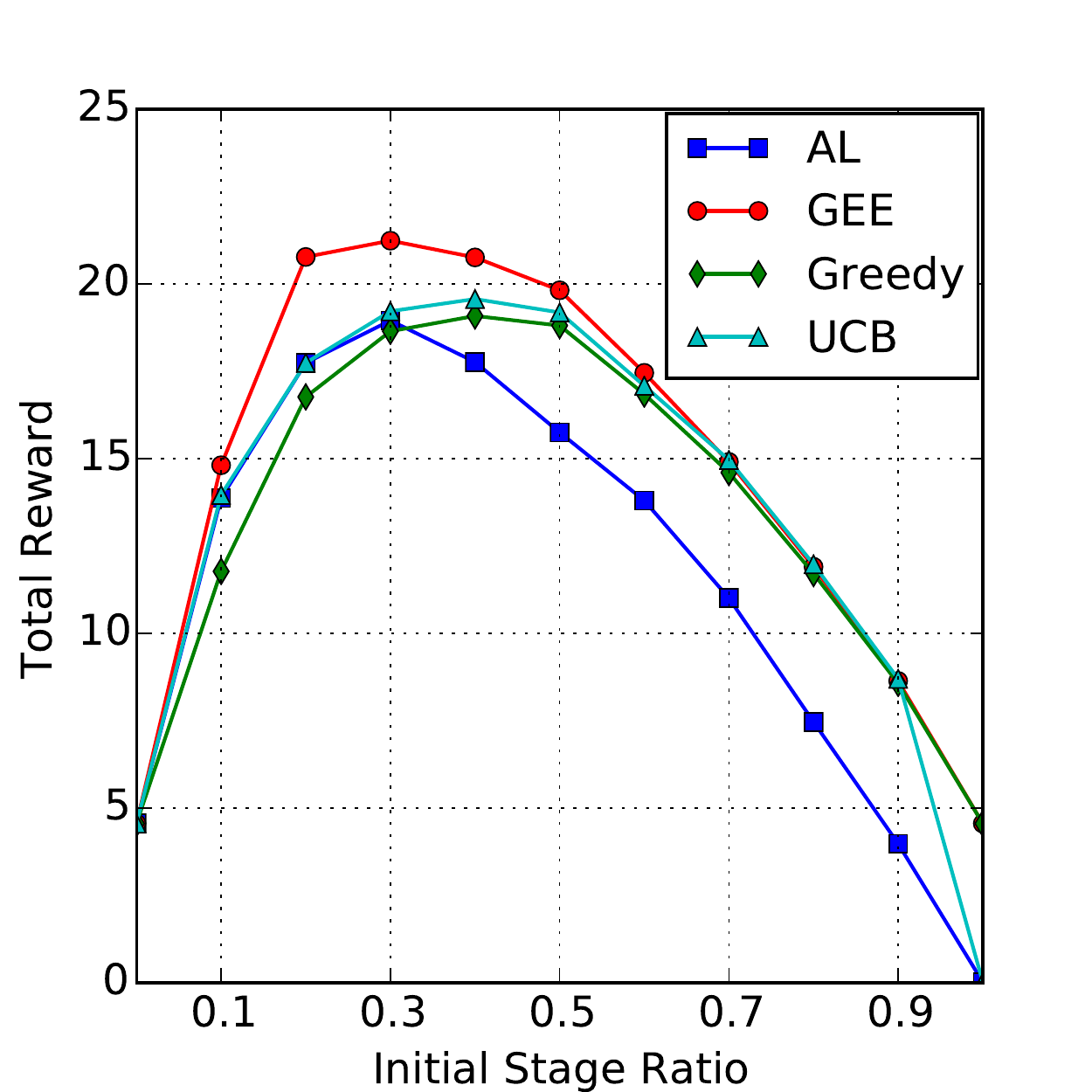}
	\label{fig:m+n=10}
	}
\subfigure[$N = 20$]{
	\includegraphics[width=1.8in]{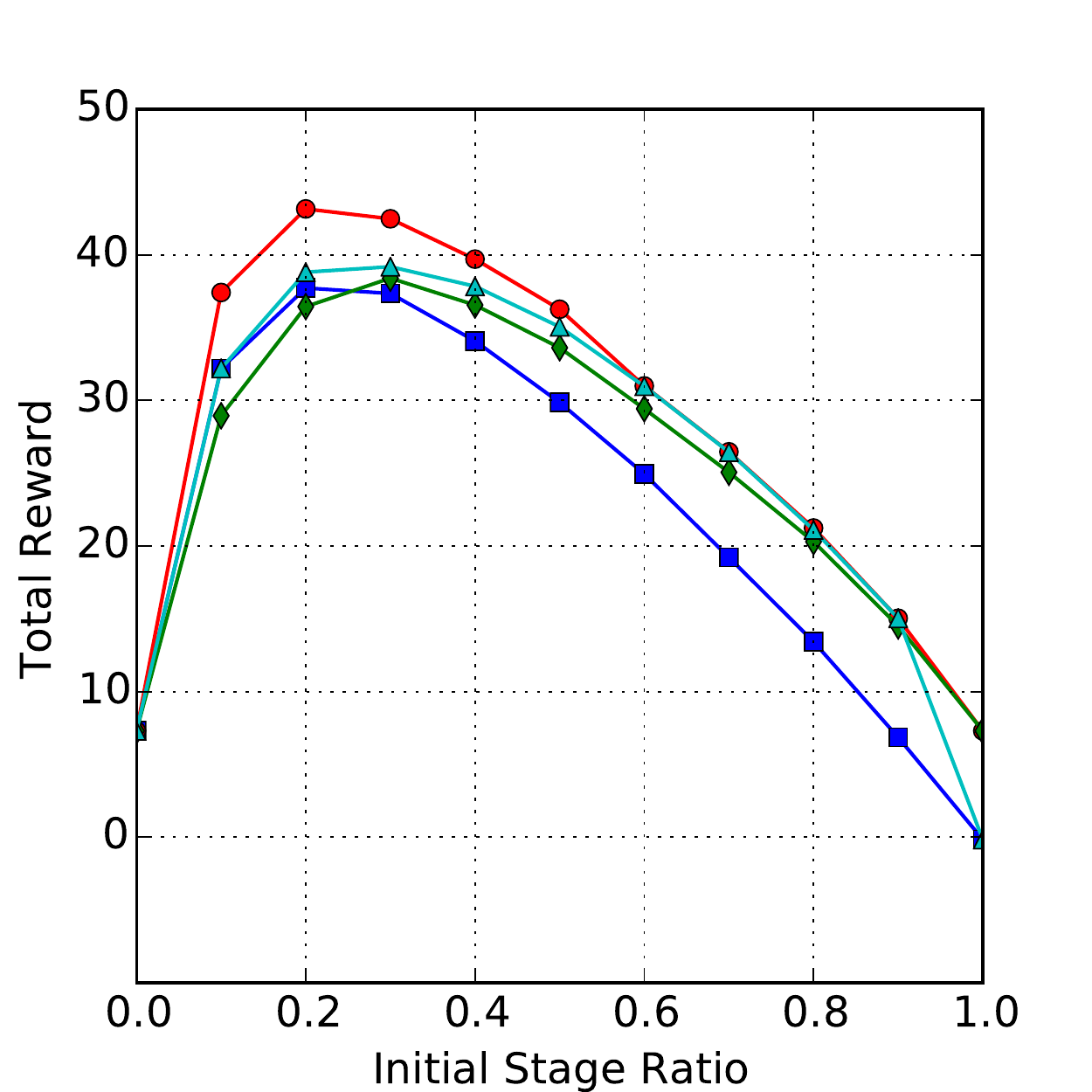}
	\label{fig:m+n=20}
}\\
\subfigure[$N = 30$]{
	\includegraphics[width=1.8in]{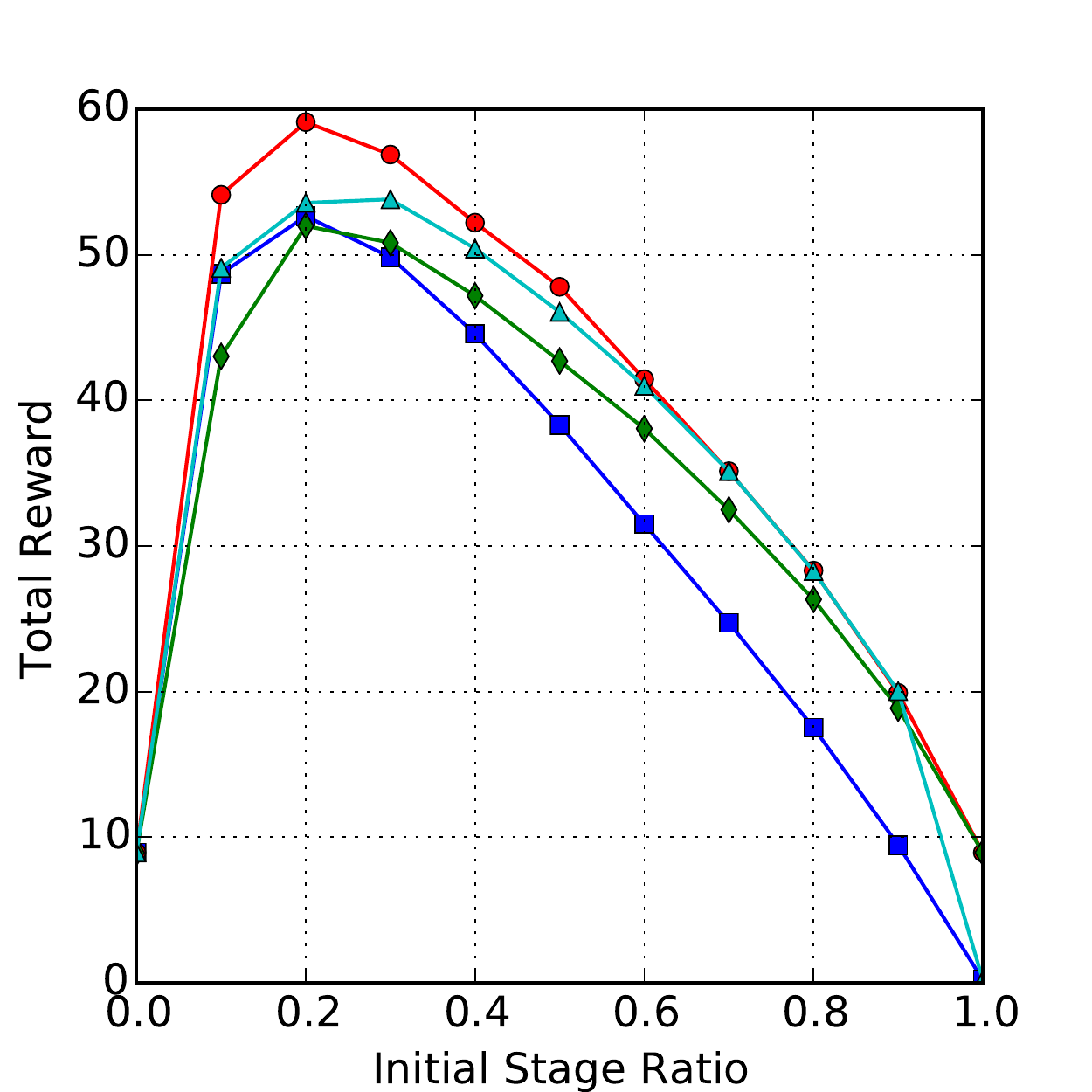}
	\label{fig:m+n=30}
}
\subfigure[$N= 40$]{
	\includegraphics[width=1.8in]{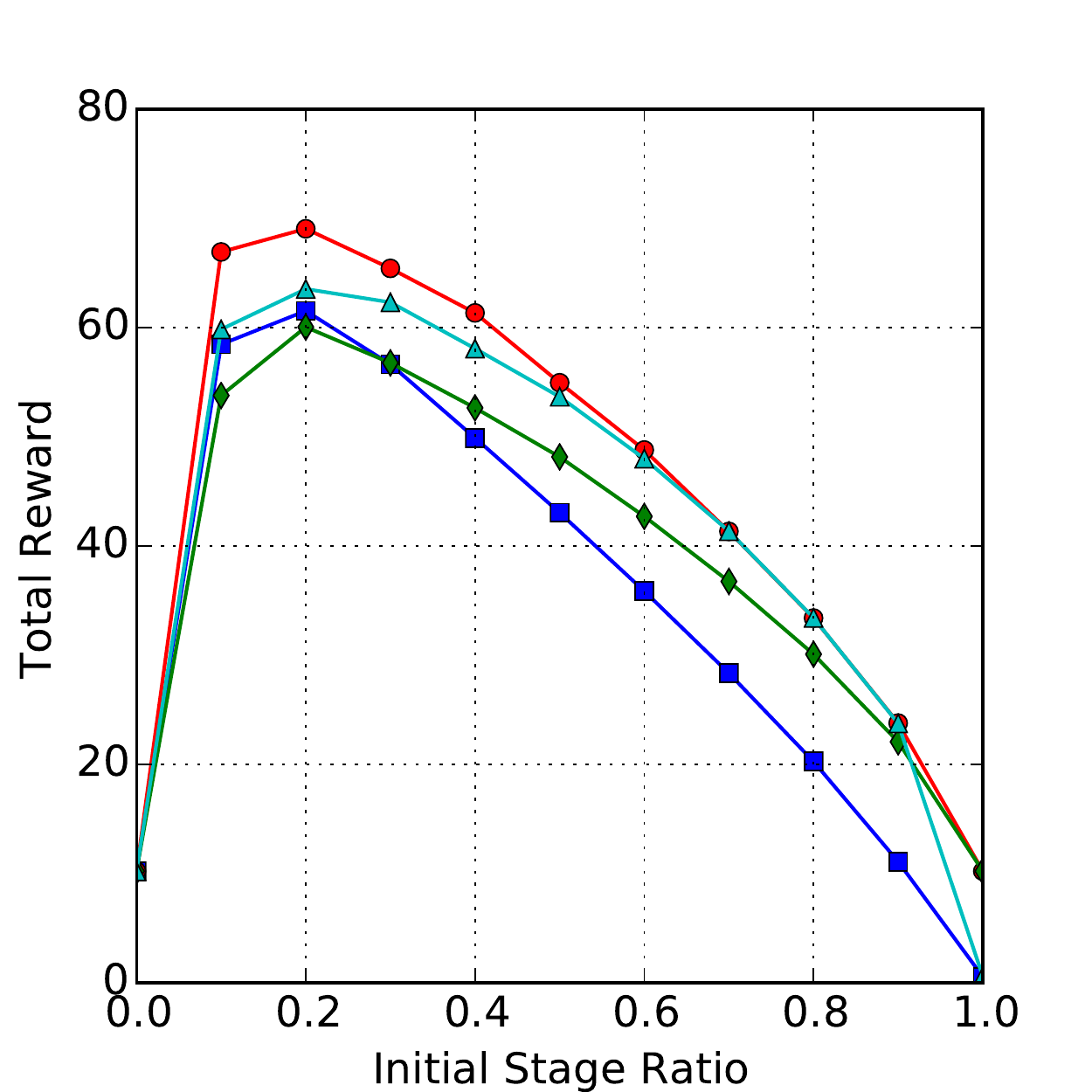}
	\label{fig:m+n=40}
}
\captionsetup{width=0.85\textwidth}
\caption{Total reward comparison of different algorithms on the synthetic data. The $x$-axis is $m/N$, the ratio of users to choose at the initial stage, and the $y$-axis is the total reward of both stages.}
\label{fig:m+n}
\end{figure}

\begin{table}[b]
\centering
\small

\caption{Total reward compared using synthetic data.}

\label{tab:compare}
\begin{tabular}{c|cccc}
\hline
Algorithm & $N = 10$ & $N = 20$ & $N = 30$ & $N = 40$ \\ \hline

Greedy & 19.084 & 38.919 & 52.517 & 60.55\\
AL & 18.953 & 37.719 & 52.655 & 62.537\\
UCB & 19.568 & 39.903 & 54.632 & 63.959\\
GEE & 21.238 & 43.151 & 59.315 & 69.198\\ \hline
\textbf{Improvement} & 8.5\%& 8.1\%& 8.6\%& 8.2\%\\ \hline

    \end{tabular}
\end{table}

\subsubsection{\textbf{Compared Methods}}
 We compare our proposed \textsf{GEE} algorithm to the following algorithms. (i) \textsf{Greedy}. \textsf{Greedy} method chooses the initial-stage users with the highest expected feedback. (ii) Active learning (\textsf{AL}). \textsf{AL} method chooses the users to minimise the uncertainty in the model, so that the users with the highest variances are chosen \cite{harpale2008personalized,rubens2011active}. (iii) Upper confidence bound (\textsf{UCB}). \textsf{UCB} method chooses the initial-stage users with the highest values calculated as the linear combination of the expected reward and the standard deviation \cite{srinivas2009gaussian}. All the algorithms select the second-stage users greedily after the system's state is updated with observations.

\begin{figure}[t]
\centering
	\subfigure[$N =10$]{
	\includegraphics[width=1.8in]{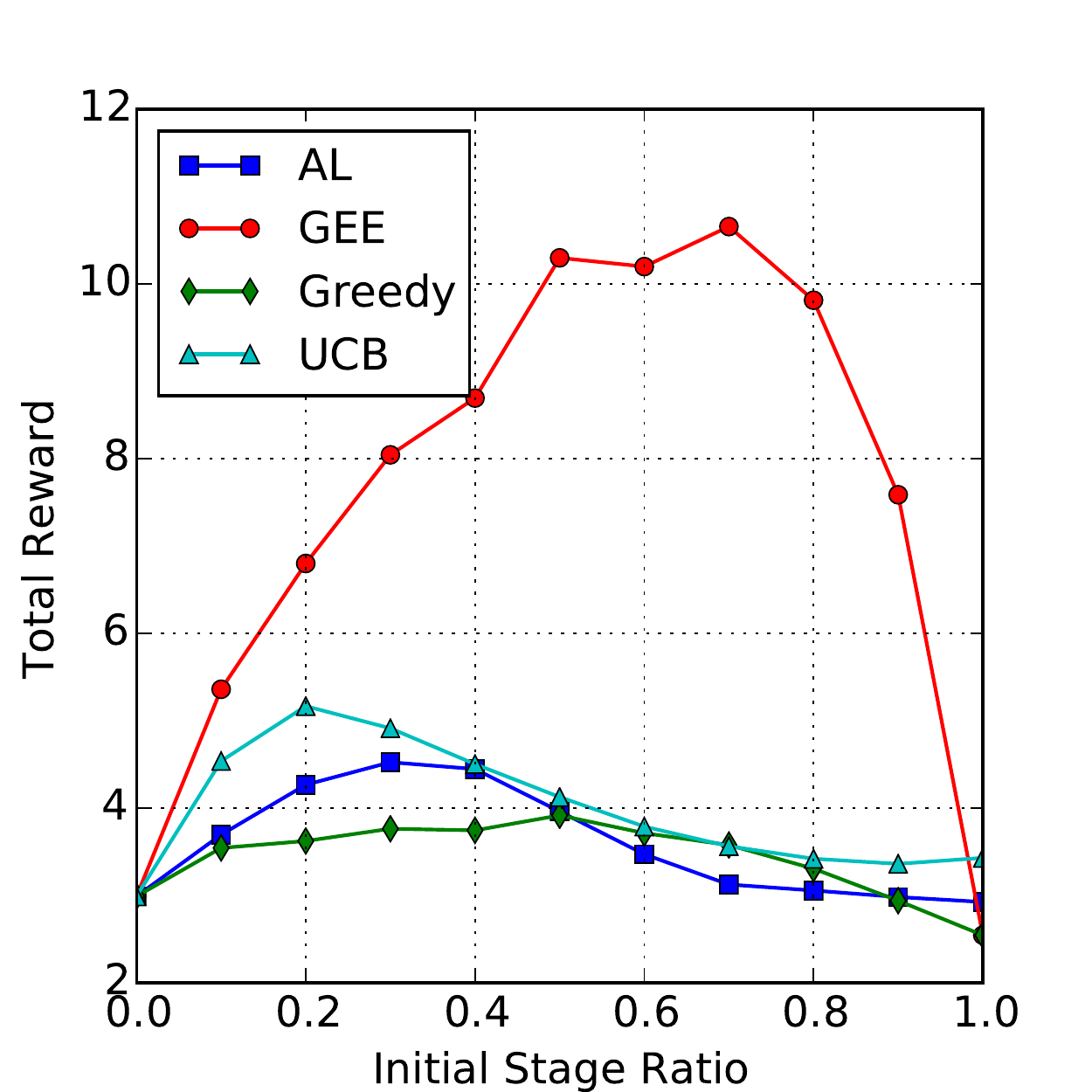}
	\label{fig:ml100k-m+n=10}
	}
\subfigure[$N = 20$]{
	\includegraphics[width=1.8in]{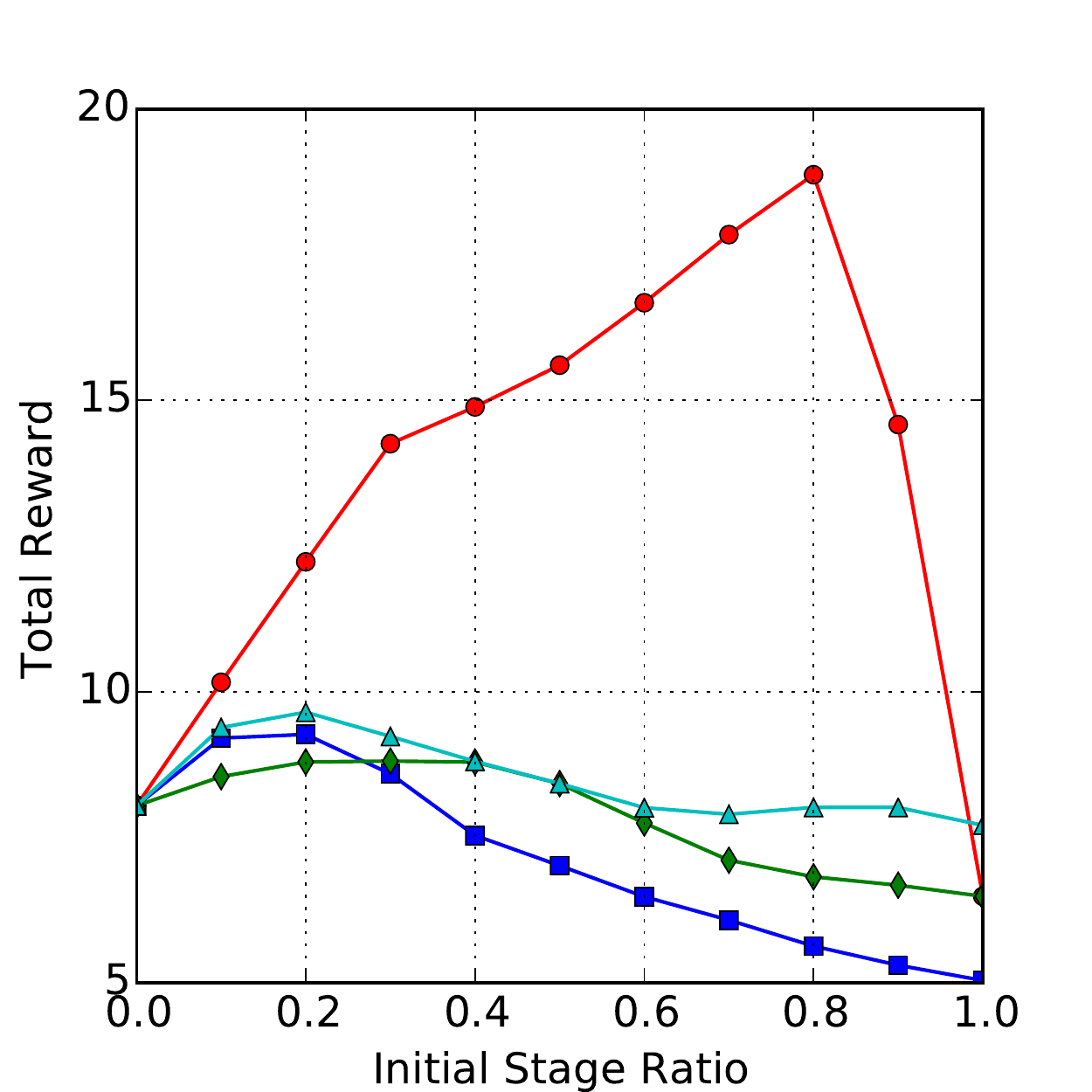}
	\label{fig:ml100k-m+n=20}
}\\
\subfigure[$N = 40$]{
	\includegraphics[width=1.8in]{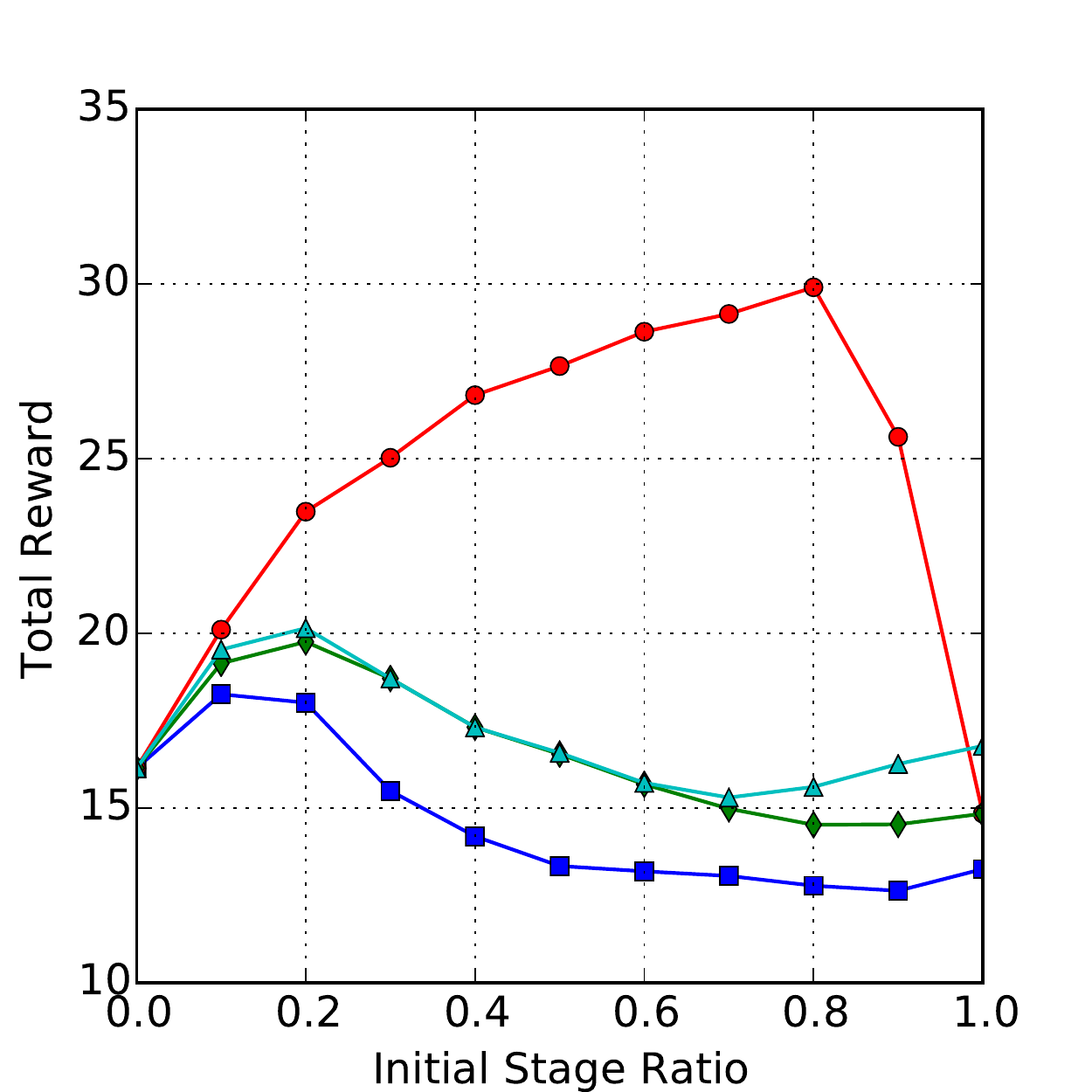}
	\label{fig:ml100k-m+n=40}
}
\subfigure[$N = 80$]{
	\includegraphics[width=1.8in]{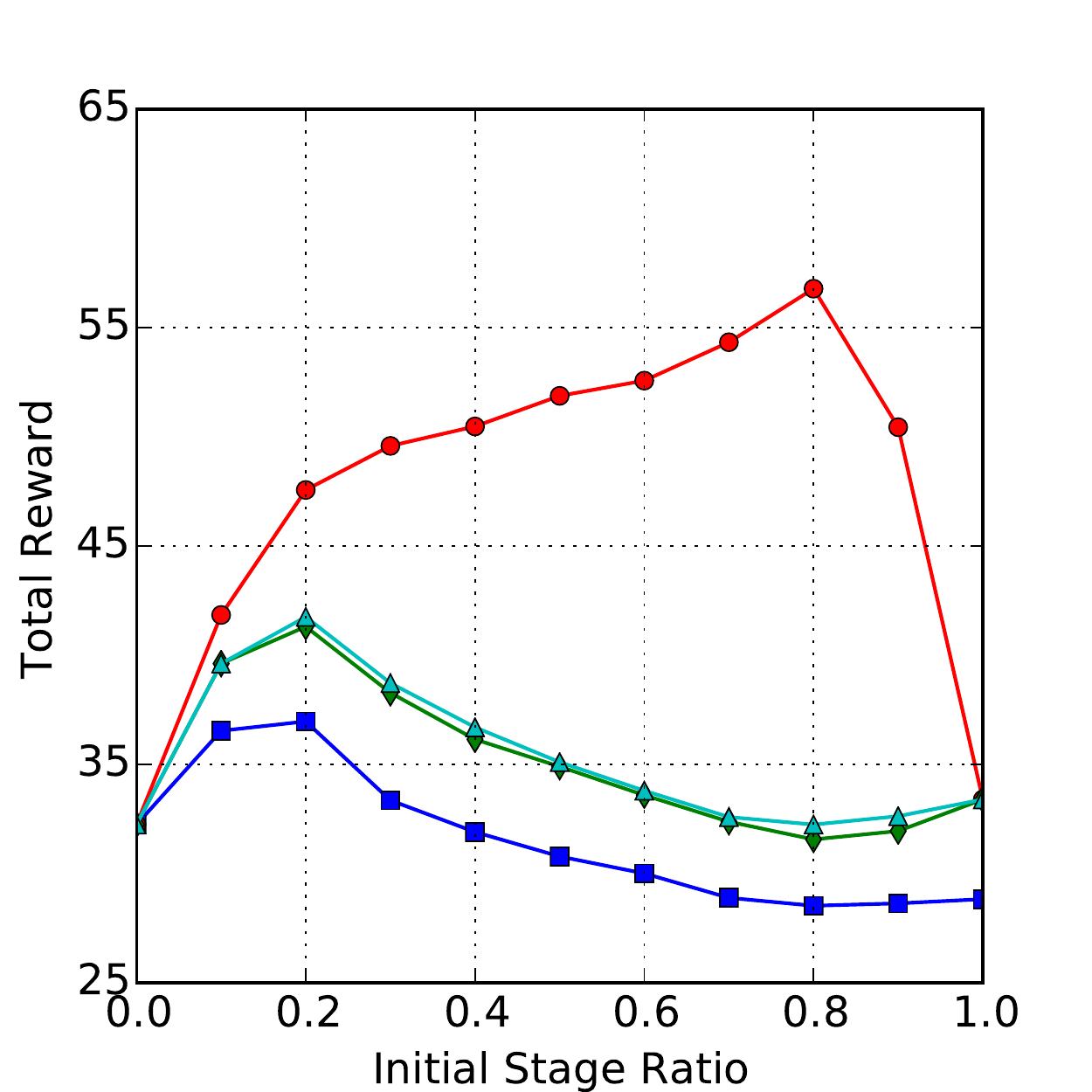}
	\label{fig:ml100k-m+n=80}
}
\captionsetup{width=0.85\textwidth}
\caption{Total reward comparison of different algorithms on the MovieLens 100K data. The $x$-axis is $m/N$, the ratio of users to choose at the initial stage, and the $y$-axis is the total reward of both stages.}
\label{fig:ml100k-m+n}
\vspace{5pt}
\end{figure}

\begin{table}[t]
\centering
\small
\caption{Total reward compared on MovieLens.}

\label{tab:ml100k-compare}
\begin{tabular}{ c|cccc }
\hline
Algorithm & $N = 10$ & $N = 20$ & $N = 40$ & $N = 80$ \\ \hline
Greedy & 4.255 & 8.95 & 20.75 & 45.26\\
AL & 4.705 & 9.91 & 21.715 & 41.665\\
UCB & 5.38 & 10.2 & 21.715 & 45.26\\
GEE & 12.125 & 19.48 & 31.05 & 60.97\\\hline
\textbf{Improvement} & 125.4\%& 91.0\%& 43.0\%& 34.7\%\\\hline
 \end{tabular}
\vspace{18pt}
\caption{Total hit number compared on MovieLens.}
\label{tab:ml100k-compare-hit}
\begin{tabular}{ c|cccc }
\hline
Algorithm & $N = 10$ & $N = 20$ & $N = 40$ & $N = 80$ \\ \hline
Greedy & 0.845 & 1.745 & 4.045 & 8.73\\
AL & 0.875 & 1.905 & 4.155 & 7.815\\
UCB & 1.015 & 1.955 & 4.155 & 8.73\\
GEE & 2.245 & 3.245 & 5.325 & 10.225\\\hline
\textbf{Improvement} & 121.2\%& 66.0\%& 28.2\%& 17.1\%\\\hline
 \end{tabular}
\end{table}

\subsubsection{\textbf{Results}}
The results are shown in Figure~\ref{fig:m+n}, with the evaluation measure as the total reward gained from the two stages. The result of the original GEE algorithm (Eq. (\ref{eq:mf_gee})) is shown and we emphasise that the result of the GEE algorithm with the intra-stage independence assumption produces similar results.

From this figure, we can make the following observations. (i) For all the algorithms, the performance improves as $m$ increases. This shows that by separating the recommendation process into two stages the performance can be greatly improved over a PRP-like  once-for-all batch solution. (ii) For all the algorithms, the total reward
increases more sharply than it drops after the performance peak. This phenomenon indicates that a small portion of allocation of users in the initial stage can significantly improve the overall performance. Note that in our synthetic data generation, we have used $K$=5, and the peak is also around $m =5$. Therefore, the dimension of the latent factor model may be an indicator of the allocation ratio. The best result gained with optimal parameters of each algorithm is shown in Table \ref{tab:compare}.
\vspace{-2pt}

\subsection{Experiments on the MovieLens Dataset}
\label{sec:movielens}

\subsubsection{\textbf{Experiment setup}}
As our study is a theoretical one, we use a relatively small research-based dataset MovieLens
100K, which is relatively small, containing 943 users and 1,682 movies, with altogether 100,000 ratings ranging from 1 to 5.
To
conduct the experiment, we first divide the dataset into the training
set and test set. For the sake of simulating cold-start item
recommendations, we first randomly choose $200$ items with sufficient
numbers of ratings (at least $50$) as the test cold-start items, and
use their ratings as the groundtruth in the test dataset. The ratings
between users and the remaining items are used to train the
model. Similar to the synthetic data experiment, we compare our
algorithms with \textsf{Greedy}, \textsf{AL} and \textsf{UCB}. After
observing the feedback, the system updates according to the user-based
CF model suggested by Eq. (\ref{eq:update_mu_approx}). The results are evaluated by using
both the total reward, and the total hit number -- the total number of ratings equal or above 4 of the two stages.
To be consistent
with what the user-based CF model suggests, we use the independent intra-user
assumption for the GEE algorithm used.

\subsubsection{\textbf{Results}}

The results are shown in Figure~\ref{fig:ml100k-m+n}, and Tables
\ref{tab:ml100k-compare} and \ref{tab:ml100k-compare-hit} with
$N = 10, 20, 40$ and $80$ respectively. Both the total reward and
the total hit number measures are compared. Here the total hit number is
defined as the total number of ratings collected which are 4 or above. We can see significant improvements over all
four cases with the implementation of our algorithm. Similar to the
synthetic experiment results, all algorithms show a peaking manner as
$m$ increases. From Tables \ref{tab:ml100k-compare} and \ref{tab:ml100k-compare-hit}
we can see that the improvements evaluated by using the total reward are even
higher than the total hit number, which may be the result of targeting
directly to the optimal reward in our objective function.

\section{Conclusion and Future Work}
\label{sec:con}
In this paper, we presented a novel two-stage recommendation process to
address the cold-start problems, with an item cold-start problem as a working example. We formulated the problem using both a
correlated-user model and a matrix factorisation model with POMDP.
With the exact solution suggested by value iteration, we concluded that
the users to choose at the initial stage should be not only of high expected values,
but also highly correlated with potential users in the next stage -- a property that can guide the system to find promising users in the next stage. With this
finding, we proposed the approximate algorithm guided exploitation-exploration (GEE). We conducted initial experiments using GEE and compared the results with
several baseline algorithms on both a synthetic and a real dataset, which confirmed
the effectiveness of our algorithm.

\vspace{10pt}
For future work, we plan to extend the two-stage process to multiple
stages and conduct larger scale experiments to study the scalability.
We are also interested in obtaining the optimal trade-off parameter
$\lambda$ and the ratio of exploitation-exploration $m/n$
theoretically.
\vspace{-2pt}

{
\bibliographystyle{abbrv}
\bibliography{ictir15}
}

\end{document}